\documentclass[referee]{aa}  

\usepackage{graphicx}
\usepackage{txfonts}

\usepackage{subfigure}
\usepackage{color}           
\usepackage{gensymb}           

\begin{document} 
	

   \title{Energy estimation of small-scale jets from the quiet-Sun region}

   \subtitle{}

   \author {Fanpeng Shi   \inst{1,2} 
          \and
          Dong Li  \inst{1,2}                     
          \and
          Zongjun Ning  \inst{1,2}       
          \and
          Jun Xu  \inst{1,2}                      
          \and
          Yuxiang Song  \inst{1,2}      
          \and
          Yuzhi Yang  \inst{1,2}          
          }

   \institute{Key Laboratory of Dark Matter and Space Science, Purple Mountain Observatory, Chinese Academy of Sciences, Nanjing 210023, China\\
              \email{shifp@pmo.ac.cn}
         \and
             School of Astronomy and Space Science, University of Science and Technology of China, Hefei 230026, China\\
             }

   \date{ }


  \abstract
   {Solar jets play a role in the coronal heating and the supply of solar wind.}
   {This study calculated the energies of 23 small-scale jets emerging from a quiet-Sun region to investigate their contributions for coronal heating.}
   {We used the data from the the High Resolution Imager (HRI) of the Extreme Ultraviolet Imager (EUI) on board the Solar Orbiter (SolO). Small-scale jets were observed by the HRI$\mathrm{_{EUV}}$ $ 174 $ \AA\ passband in the high cadence of $ 6 $ s. These events were identified by the time-distance stacks along the trajectories of jets. Using the simultaneous observation from the Atmospheric Imaging Assembly (AIA) on board the Solar Dynamics Observatory (SDO), we also performed the differential emission measure (DEM) analysis on these small-scale jets to obtain the physical parameters of plasma, which enabled the estimation of the kinetic and thermal energies of jets.}
   {We found that most of jets exhibited the common unidirectional or bidirectional motions, while some showed the more complex behaviors, namely the mixture of unidirection and bidirection. A majority of jets also presented the repeated eruption blobs (plasmoids), which may be the signatures of the quasi-periodic magnetic reconnection that have been observed in the solar flares. The inverted Y-shaped structure can be recognized in several jets. These small-scale jets typically have the width of $\sim$$0.3$ Mm, the temperature of $\sim$$1.7$ MK, the electron number density of $\gtrsim$$10^{9}\ \mathrm{cm^{-3}}$, with speeds in a wide range of $\sim$$20$--$170$ $\mathrm{km\ s^{-1}}$. Most of these jets have the energy of $10^{23}$--$10^{24}\ \mathrm{erg}$, which is marginally smaller than the energy of typical nanoflares. The thermal energy fluxes of 23 jets are estimated to be (0.74--2.96) $\times 10^{5}\ \mathrm{erg\ cm^{-2}\ s^{-1}}$, which is almost on the same order of magnitude as the energy flow required for heating the quiet-Sun corona, albeit the kinetic energy fluxes vary over a large range because of their strong dependence on velocity. Furthermore, the frequency distribution of thermal energy and kinetic energy both follow the power-law distribution $N(E) \propto$ $E^{-\alpha}$.}
   {Our observations suggest that although these jets cannot provide sufficient energy for the heating of the whole quiet-Sun coronal region, they are likely to account for a significant portion of the energy demand in the local regions where the jets occur.}


   \maketitle

%

\section{Introduction}\label{intro}

Solar jets are collimated beam-like plasma ejections in the solar atmosphere. They are prevalent and can occur at different scales, and are considered as the important source of coronal heating and the solar wind. \citep{Raouafi2016SSRv..201....1R, Shen2021RSPSA.47700217S}. Solar jets also represent the characteristics of magnetic reconnection, with bidirectional jets strongly supporting this explanation due to their double outflows \citep{Innes1997Natur.386..811I, Zheng2018ApJ...861..108Z, Shen2019ApJ...883..104S}. Over the decades, thanks to the improvement of the spatiotemporal resolution of observation, small-scale jets in different layers of the solar atmosphere have been studied in detail. \cite{Shibata2007Sci...318.1591S} investigated the inverted Y-shaped chromospheric anemone jets in active regions, suggesting that the heating of the chromosphere and corona may be related to small-scale ubiquitous reconnection. \cite{Ji2012ApJ...750L..25J} reported the ultrafine channels for coronal heating, in which continuous small-scale upward energy flows originate from photosphere and subsequently light up the corona. \cite{Tian2014Sci...346A.315T} observed the prevalent small-scale jets from the networks of the solar transition region and chromosphere, with speeds on the order of $\sim$$100\ \mathrm{km\ s^{-1}}$ and temperatures of $>$$10^{5}$ K, which are likely the source for the solar wind. There are also some other detailed studies, such as the chromospheric small-scale jets \citep{Wang2021ApJ...913...59W, Hong2022ApJ...928..153H}, chromospheric spicules \citep{Samanta2019Sci...366..890S}, recurrent jets from sunspot light bridges \citep{Tian2018ApJ...854...92T}, small-scale IRIS jets \citep{Li2018MNRAS.479.2382L}, and coronal nanojets \citep{Antolin2021NatAs...5...54A}.

The high spatiotemporal resolution coronal extreme ultraviolet (EUV) images observed by the Hi-C 2.1, provided new insights into the study of small-scale jets owing to the discoveries of dot-like, loop-like, and surge/jet-like EUV brightenings at small spatial scales \citep{Tiwari2019ApJ...887...56T, Panesar2019ApJ...887L...8P}. Recently, the observations by the Extreme Ultraviolet Imager (EUI; \citealt{Rochus2020A&A...642A...8R}) on board the Solar Orbiter (SolO; \citealt{Muller2020A&A...642A...1M}), have revealed the transient small-scale brightenings in the quiet solar corona termed campfires \citep{Berghmans2021A&A...656L...4B, Chen2021A&A...656L...7C, Panesar2021ApJ...921L..20P}, the moving structures in ultraviolet bright points \citep{Li2022A&A...662A...7L}, the persistent null-point reconnection \citep{Cheng2023NatCo..14.2107C}, as well as the small-scale plasma flow and jets \citep{Hou2021ApJ...918L..20H, Chitta2021A&A...656L..13C, Mandal2022A&A...664A..28M, Chitta2023Sci...381..867C}.

The energy of a small-scale jet usually has the order of magnitude of nanoflare-like ($\sim$$10^{24}\ \mathrm{erg}$, \citealt{Parker1988ApJ...330..474P}). \cite{Hou2021ApJ...918L..20H} estimated the energy of a typical coronal microjets, with thermal energy of $\sim$$3.9 \times\ 10^{24}\ \mathrm{erg}$ and kinetic energy of $\sim$$2.9 \times\ 10^{23}\ \mathrm{erg}$. \cite{Antolin2021NatAs...5...54A} estimated the total energy ("thermal + kinetic") released by a coronal nanojet to be $(7.8$--$17.3) \times 10^{24}\ \mathrm{erg}$, and they expected that this range of energies corresponds to the high end of the nanojet distribution, most of which is likely to be unresolved by observations. Additionally, the small-scale EUV brightening or bursts also have the energy of $10^{24}\ \mathrm{erg}$ per event \citep{Chitta2021A&A...647A.159C}. \cite{Li2018MNRAS.479.2382L} estimated that an upper limit of the thermal energy of the coronal EUV brightening caused by the IRIS jets is around $7 \times 10^{24}\ \mathrm{erg}$.
 Using the observation of EUV brightening, \cite{Purkhart2022A&A...661A.149P} studied the nanoflare energy distributions in quiet-Sun regions, and found that the mean energy flux of $(3.7\pm$$1.6) \times 10^{4}\ \mathrm{erg\ cm^{-2}\ s^{-1}}$ is one order of magnitude smaller than the coronal heating requirement. They also found that clusters of high energy flux (up to $3 \times 10^{5}\ \mathrm{erg\ cm^{-2}\ s^{-1}}$) are surrounded by extended regions with lower activity.

In this study, we investigated 23 small-scale jets emerging from a quiet-Sun region and estimated their energies. Several events showed the ambiguous inverted Y-shaped structures resembling the standard or blowout jets as described by \cite{Moore2010ApJ...720..757M} and \cite{Sterling2015Natur.523..437S}. Most of them exhibited the nearly collimated simple structures similar to the coronal microjets \citep{Hou2021ApJ...918L..20H} and picoflare jets \citep{Chitta2023Sci...381..867C}. This paper is organized as follows: Section \ref{obs} introduces the observations and data analysis. Section \ref{data} analyzes three examples of small-scale jets. Section \ref{discuss} is the discussion and conclusions.

\section{Observations and Data Analysis} \label{obs}
On 2023 March 30, the observation of the 174 \AA\ EUV High Resolution Imager (HRI$\mathrm{_{EUV}}$) of SolO/EUI started at 15:00:15 UT and ended at 16:07:09 UT (Earth time) in a cadence of 6 s (a total of 670 frames). The spacecraft was at a distance of $\sim$0.379 AU from the Sun, and its position respect to solar longitude was $\sim$$6.2^{\circ}$ west from Sun-Earth line. The HRI$\mathrm{_{EUV}}$ 174 \AA\ plate scale is 0.492 arcsec per pixel, corresponding to a linear scale of $\sim$134 km per pixel at this distance. This passband is sensitive to emissions from plasma at temperatures of roughly 1 MK. The level-2 calibrated data \citep{Kraaikamp_euidatarelease6}\footnote{https://www.sidc.be/EUI/data/releases/202301\_release\_6.0/release\_notes.html} were used and the remaining jitter in image sequence was removed using the cross-correlation method.

\begin{figure*}
	\centering
	\includegraphics[width=16 cm]{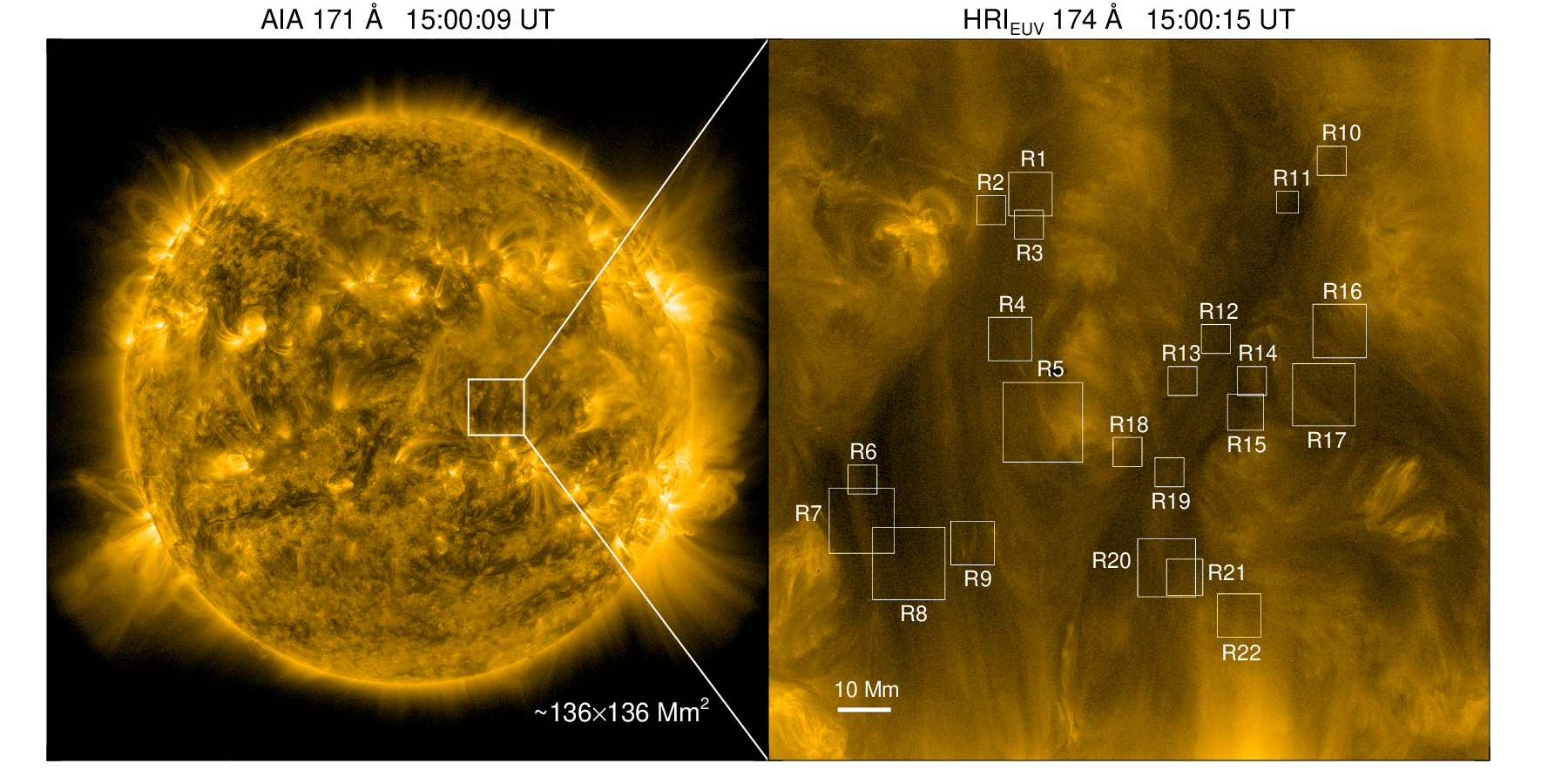}	
	\caption{AIA $171$ \AA\ image of the solar disk (left), and the white box represents the selected HRI$_\mathrm{{EUV}}$ $174$ \AA\ FOV ($\sim$$136 \times 136\ \mathrm{Mm^{2}}$). Twenty two regions (R1--R22) analyzed in this study are labeled in the HRI$_\mathrm{{EUV}}$ $174$ \AA\ image (right).}
	\label{figure1}%
\end{figure*}

The HRI$\mathrm{_{EUV}}$ 174 \AA\ had a field-of-view (FOV) at a quiet-Sun region on the north side of the NOAA active region (AR) 13262, as outlined by the white box in Fig.~\ref{figure1} (left panel). HRI$\mathrm{_{EUV}}$ 174 \AA\ observation is shown in the movie (movie\_hri174.mp4). During this observation period, there were plenty of jet eruptions in this area. A total of twenty two regions were detected and marked by R1--R22 in Fig.~\ref{figure1} (right panel). Some jets ejected matter over long distances, while others ejected short distances. We used the white boxes with different scales to cover the jets.

Six EUV channels of the Atmospheric Imaging Assembly (AIA; \citealt{Lemen2012SoPh..275...17L}) and the line-of-sight (LOS) magnetograms of Helioseismic and Magnetic Imager (HMI; \citealt{Scherrer2012SoPh..275..207S}) on board the Solar Dynamics Observatory (SDO; \citealt{Pesnell2012SoPh..275....3P}) were also used in this study. AIA 94 \AA\ ($T$$\sim$$6$ MK), 131 \AA\ ($T$$\sim$$10$ MK), 171 \AA\ ($T$$\sim$$0.7$ MK), 193 \AA\ ($T$$\sim$$1.6$ MK), 211 \AA\ ($T$$\sim$$2$ MK), and 335 \AA\ ($T$$\sim$$2.5$ MK), mainly reflect the emissions from the corona. The response of AIA 171 \AA\ passband is similar to the HRI$\mathrm{_{EUV}}$ 174 \AA. These EUV passbands have a cadence of 12 s and a plate scale of 0.6 arcsec ($\sim$432 km). The HMI LOS magnetograms have a cadence of 45 s and a plate scale of 0.5 arcsec ($\sim$360 km). All of the above data have been removed solar rotation by derotating to 15:00:09 UT as seen in Fig.~\ref{figure1}. The differential emission measure (DEM) analysis \citep{Cheung2015ApJ...807..143C, Su2018ApJ...856L..17S} was performed on co-aligned AIA images to extract the physical parameters of plasma. In this procedure, we averaged the AIA intensities over two frames, i.e., the reconstructed AIA image sequences have a cadence of 24 s. The temperature range of inversion is set as $5.5 \leq \log_{10}T \leq 7.6$, following with previous suggestions \citep{Hannah2012A&A...539A.146H, Cheung2015ApJ...807..143C, Su2018ApJ...856L..17S}.

The DEM function is an instrument-independent function that characterizes the electron and temperature of optically thin plasma in thermal equilibrium. The DEM function (in units of $\mathrm{cm^{-5}\ K^{-1}}$) is defined as the squared electron density $n_{\mathrm{e}}^2$ integrated along the LOS depth $z$,
\begin{equation}
	\mathrm{DEM}(T) = n_{\mathrm{e}}^2\frac{dz}{dT},
\end{equation}
and the emission measure (EM)-weighted temperature is,
\begin{equation}
		\bar{T} = \frac{\int_{}^{}\mathrm{DEM}(T)\times TdT}{\int_{}^{}\mathrm{DEM}(T)dT}= \frac{\int_{}^{}\mathrm{DEM}(T)\times TdT}{\mathrm{EM}},
\end{equation}
the $ \mathrm{DEM}(T)$ also allow us to estimate the unknown electron density $n_{\mathrm{e}}$ via the measurement of $z$,
\begin{equation}\label{ne}
	\begin{array}{l}
	\mathrm{EM} = \int_{}^{}\mathrm{DEM}(T)dT =  \int_{}^{}n_{\mathrm{e}}^2dz = n_{\mathrm{e}}^2z,
	\end{array}
\end{equation}
where $n_{\mathrm{e}}$ defines a mean electron density that is averaged over the LOS depth $z$ with a filling factor of unity \citep{Aschwanden2000ApJ...535.1047A, Aschwanden2015ApJ...802...53A}.

\begin{figure*}
	\centering
	\includegraphics[width=16 cm]{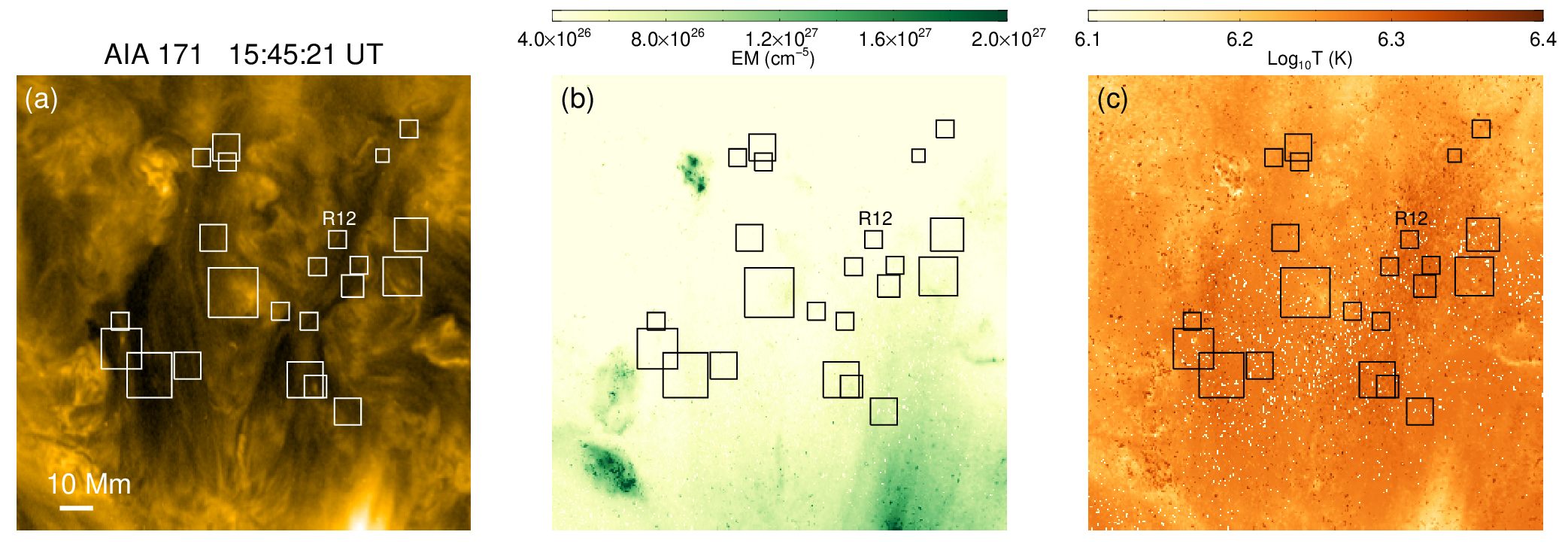}
	\caption{ (a)--(c) AIA 171 \AA\ map, the EM map, and the EM-weighted temperature map, respectively. The FOV of map covers the R1--R22 regions.}
	\label{figure2}%
\end{figure*}

Fig.~\ref{figure2}(a) shows the AIA 171 \AA\ map taken at 15:45:21 UT, from which we can see the similar structure as the HRI$_\mathrm{{EUV}}$ $174$ \AA\ in Fig.~\ref{figure1}. Based on the DEM analysis in this region, Fig.~\ref{figure2}(b)--(c) give the corresponding EM and EM-weighted temperature maps. Twenty two regions are marked with small boxes.

We describe here the assumptions and formulae used to calculate the jet energies. If we assume that these small-scale jets are elongated cylinders and our LOS is perpendicular to the direction of jet motion, then the depth $z$ is equal to the jet width ($w$). Assuming that the structure of jet is isothermal and homogeneous (with a constant electron density), the total thermal energy can be expressed as (e.g., \citealt{Aschwanden2015ApJ...802...53A}),
	\begin{equation}\label{thermal}
		E_{\mathrm{th}} =   3n_\mathrm{e}k_\mathrm{B}T_\mathrm{e}V_\mathrm{{jet}},
	\end{equation}
	where $n_\mathrm{e}=\sqrt{\dfrac{\Delta \mathrm{EM}}{w}}$ is the jet density after considering the variation in EM relative to the pre-event value ($\Delta \mathrm{EM}$ is the increase in EM, e.g., \citealt{Purkhart2022A&A...661A.149P, Long2023ApJ...944...19L}), $k_\mathrm{B} = 1.38 \times 10^{-16}\ \mathrm{erg\ K^{-1}}$ is Boltzmann constant, $T_\mathrm{e}$ is temperature around the time when the EUV emission peaked, and $V_\mathrm{{jet}}=\pi (\dfrac{w}{2})^2l$ is the volume of jet. The jet width $w$ and length $l$ can be measured from HRI$_\mathrm{{EUV}}$ $174$ \AA\ image.
	
We can also calculate the mean thermal energy flux, which presumes that the total thermal energy is dissipated through the surface area of the jet throughout its entire lifetime,
	\begin{equation}
		F_{\mathrm{th}} =   \frac{3n_\mathrm{e}k_\mathrm{B}T_\mathrm{e}V_\mathrm{{jet}}}{\tau_\mathrm{{tot}}S_\mathrm{{jet}}},
	\end{equation}
	where $\tau_\mathrm{{tot}}$ is the total lifetime, $S_\mathrm{jet} \approx \pi wl\ (\pi wl \gg \pi (\dfrac{w}{2})^2)$ is the surface area of jet, then the expression of $F_{\mathrm{th}}$ can be reduced as, 
	\begin{equation}\label{thermalflux}
		F_{\mathrm{th}} =   \frac{3n_\mathrm{e}k_\mathrm{B}T_\mathrm{e}w}{4\tau_\mathrm{{tot}}},
	\end{equation}
	which is independent of the jet length.

The kinetic energy flux carried by a parcel of fluid moving with velocity $v$ is (e.g., \citealt{Chitta2023Sci...381..867C})
	\begin{equation}\label{kineticflux}
		F_{\mathrm{k}} =   \frac{1}{2}\rho v^{3},
	\end{equation}
	where $\rho \approx n_\mathrm{e}m_\mathrm{p}$ is the plasma mass density, and $m_\mathrm{p}=1.67 \times 10^{-26}\ \mathrm{g}$ is the proton mass. The $F_{\mathrm{k}}$ is the energy per unit area per unit time on a plane perpendicular to motion. Subsequently, the total kinetic energy can be derived,
	\begin{equation}\label{totalkinetic}
		E_\mathrm{k} =   \sum_{i}^{}F_{\mathrm{k}i}\tau_i S_\mathrm{bot},
	\end{equation}
	where subscript $i$ denotes the $i$th blob of jet, $\tau_i$ is its lifetime, and $S_\mathrm{bot}= \pi (\dfrac{w}{2})^2$ is bottom area of jet.

\section{Results} \label{data}

\subsection{Unidirectional jets}

\begin{figure*}
	\centering
	\includegraphics[width=16 cm]{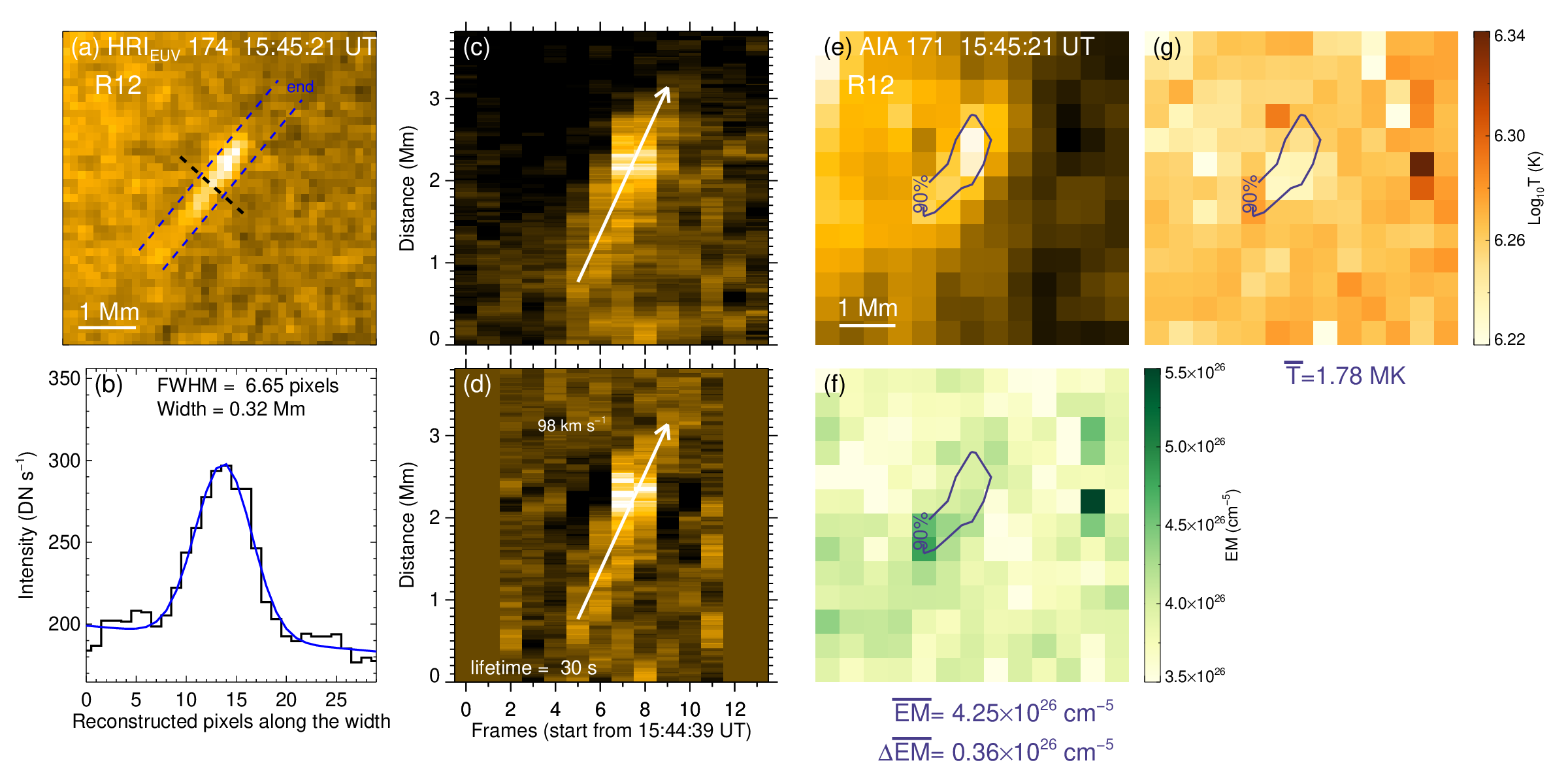}	
	\caption{(a) HRI$_\mathrm{{EUV}}$ $174$ \AA\ image of the R12 jet. (b) The estimation of jet width. (c) Time-distance diagram in the position of blue slit. (d) Smooth-subtracted image of panel (c). (e) AIA 171 \AA\ image of the R12 jet. Contour level represents the 90\% of the peak intensity of AIA 171 \AA. (f) The EM map. (g) The EM-weighted temperature map. The mean EM and temperature are averaged over the 90\% area. The $\Delta \mathrm{EM}$ is the net increase in EM relative to pre-jet value.}
	\label{figure3}%
\end{figure*}

Each blob (plasmoid) of small-scale jets recognized and analyzed in this study lasts for at least three frames (i.e., 18 s), which enabled the motion identification from HRI$_\mathrm{{EUV}}$ $174$ \AA\ images. Fig.~\ref{figure3}(a) shows the R12 jet in HRI$_\mathrm{{EUV}}$ $174$ \AA\ image, which presents a simple linear motion along a straight line. The blue slit is outlined to obtain the time-distance stacks, and the "end" represents the end of slit. The black dashed line is the position used for calculating the jet width. Fig.~\ref{figure3}(b) shows the estimation for jet width. The black profile is the reconstructed intensity along the black slit in Fig.~\ref{figure3}(a), and the blue profile represents the singe Gaussian fitting with a linear background. The FWHM of fitting result is considered as the width, which corresponds to a linear scale of $\sim$0.32 Mm. Fig.~\ref{figure3}(c) shows the time-distance diagram derived from the blue slit. To make it easier to show the variability, we also give the corresponding smooth-subtracted images for all time-distance diagrams in this study. Fig.~\ref{figure3}(d) is the smooth-subtracted image of panel (e), i.e., after subtracting the slowly-varying component, which is a 4-frame (24 s) running average of original intensity. The smoothing window is chosen individually for each event, and its size can be seen through the blank frames on both sides of the smooth-subtracted image. In Fig.~\ref{figure3}(d), there is a slanted bright stripe spanning five frames, which outlines the blob motion with velocity of $\sim$$\mathrm{98\ km\ s^{-1}}$. The jet lifetime is defined as the time interval between the beginning and end of the moving blob, i.e., 30 s.

Fig.~\ref{figure3}(e)--(f) is the zoom-in of the R12 region as labeled in Fig.~\ref{figure2}. It is obvious that the moving blob is only several bright pixels in the FOV of AIA 171 \AA. We choose the 90\% of the peak intensity as the contour of jet, which is a good representation due to its elongated structure. Therefore, the mean EM and EM-weighted temperature can be obtained within the 90\% area, their values are 4.25 $\times\ 10^{26}\ \mathrm{cm^{-5}}$ and 1.78 MK, respectively. As noted in Section \ref{obs}, the variation in EM relative to pre-jet value should be considered to estimate the jet density. The $\Delta \mathrm{EM}$ below Fig.~\ref{figure3}(f) is the net increase that is calculated over the same area as the $ \mathrm{EM}$. These EMs and EM-weighted temperatures are computed over the temperature range of $5.6 \leq \log_{10}T \leq 6.5$, where the DEM solution is well constrained. This range is almost the same as previous literature about coronal microjets \citep{Hou2021ApJ...918L..20H}. All other jets also exhibit emissions originating from a similar temperature range and lack higher temperature components, and thus we can apply the same range to EM analysis of all jets.

 Using the $\Delta \mathrm{EM}=0.36 \times 10^{26}\ \mathrm{cm^{-5}}$ and $w=0.32 $ Mm in Equation~\ref{ne}, then the $n_\mathrm{e}$ is $1.06 \times 10^{9}\ \mathrm{cm^{-3}}$, which is a reasonable value for the coronal condition. The jet length $l\sim$1.5 Mm can be roughly measured from the HRI$_\mathrm{{EUV}}$ $174$ \AA\ image. Substituting these known quantities into the Equation \ref{thermal}, the estimated total thermal energy is $9.43 \times 10^{22}\ \mathrm{erg}$, which is one order of magnitude smaller than the energy of typical nanoflares ($\sim$$10^{24}\ \mathrm{erg}$, \citealt{Parker1988ApJ...330..474P, Chitta2021A&A...647A.159C}). Thus, the thermal energy flux of R12 jet is $2.08 \times 10^{5}\ \mathrm{erg\ cm^{-2}\ s^{-1}}$, contributing $2/3$ of the canonical amount of energy required to heat the quiet-Sun corona ($3 \times 10^{5}\ \mathrm{erg\ cm^{-2}\ s^{-1}}$, \citealt{Withbroe1977ARA&A..15..363W}).

Using the Equation \ref{kineticflux} and \ref{totalkinetic}, the kinetic energy flux of R12 jet is estimated to be $8.34 \times 10^{5}\ \mathrm{erg\ cm^{-2}\ s^{-1}}$, and the kinetic energy is $2.01 \times 10^{22}\ \mathrm{erg}$. Note that the velocity calculated above is the projected velocity and may depend on the viewing angle. The kinetic energy estimation can be heavily affected by the projection effect and may not be accurate, and based on the calculation, the kinetic energy is very small compared to the thermal energy. It may not be accurate to represent the total energy of the jet simply as "$E_\mathrm{k} + E_\mathrm{th}$" and hence we do not present the total energy in this study.

\subsection{Bidirectional jets}

\begin{figure*}
	\centering
	\includegraphics[width=16 cm]{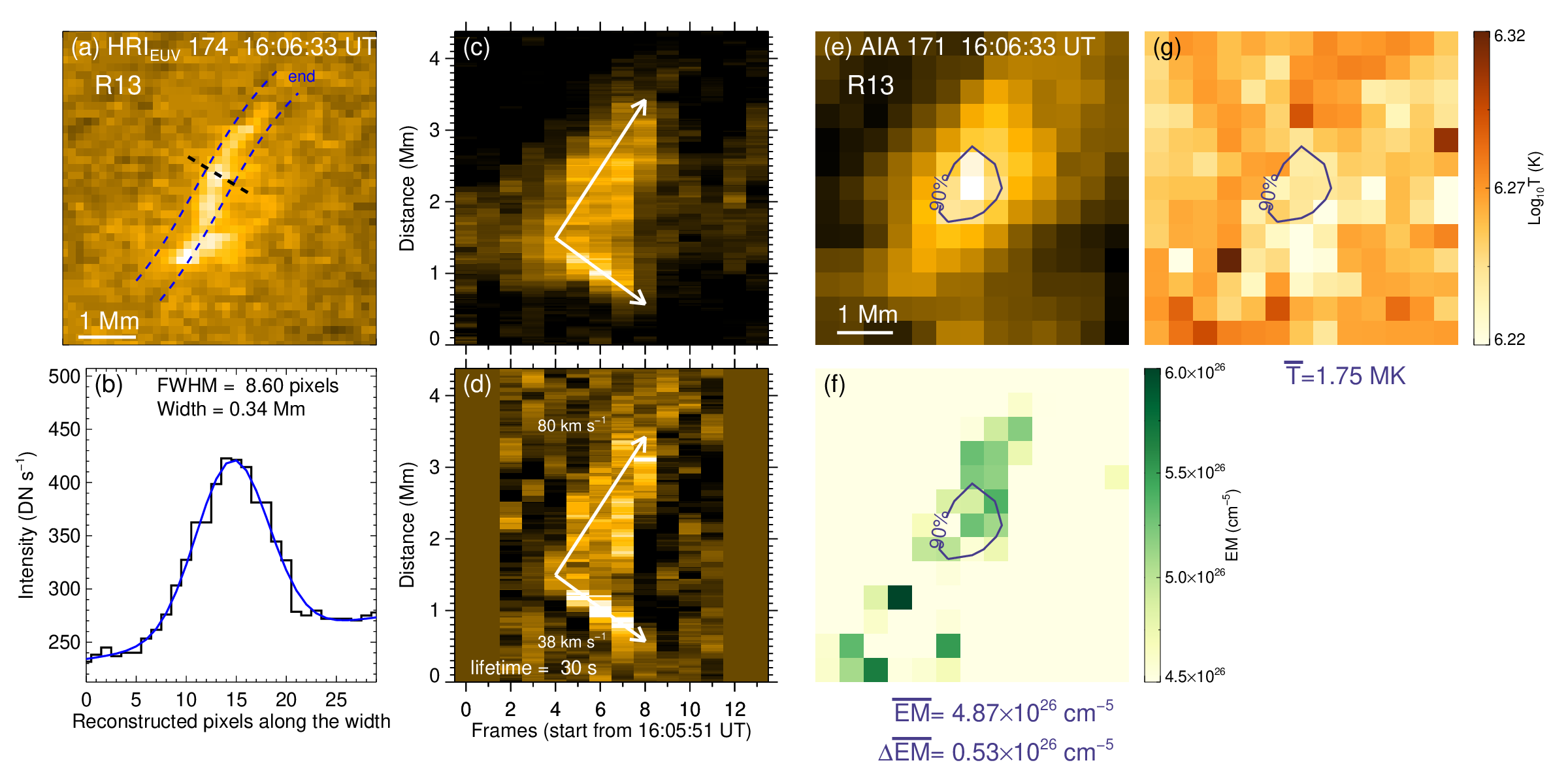}
	\caption{ (a)--(g) Similar to Fig.~\ref{figure3}, the results for R13 jets.}
	\label{figure4}%
\end{figure*}

Fig.~\ref{figure4} gives an example of bidirectional jets in R13. Similar to the analysis for R12 jet, Fig.~\ref{figure4}(a)--(d) show the HRI$_\mathrm{{EUV}}$ $174$ \AA\ image, the estimation of jet width, the time-distance diagram, and the smooth-subtracted image, respectively. In this event, two blobs have different velocities ($\sim$$80\ \mathrm{km\ s^{-1}}$ and $\sim$$38\ \mathrm{km\ s^{-1}}$) , but their lifetimes are the same as the total lifetime, i.e., 30 s. The jet length has a rough value of $\sim$3 Mm. Fig.~\ref{figure4}(e) shows the AIA 171 \AA\ image of the R13 jets, which just exhibits a few of bright pixels. The 90\% contour level is also marked. The EM and EM-weighted temperature maps are shown in Fig.~\ref{figure4}(f)--(g).

According to the above procedures, the electron density, thermal energy flux, and total thermal energy, are estimated to be $1.25 \times 10^{9}\ \mathrm{cm^{-3}}$, $2.56 \times 10^{5}\ \mathrm{erg\ cm^{-2}\ s^{-1}}$, and $2.46 \times 10^{23}\ \mathrm{erg}$, respectively. The derived thermal energy flux almost satisfies the energy demand for coronal heating in quiet Sun region. The total thermal energy is in the range of $10^{23}$--$10^{24}\ \mathrm{erg}$. As for the kinetic energy flux, two blobs should be calculated individually, they are $5.34 \times 10^{5}$ and $0.57 \times 10^{5}\ \mathrm{erg\ cm^{-2}\ s^{-1}}$. Therefore, the total kinetic energy is $1.61 \times 10^{22}\ \mathrm{erg}$, which is one order of magnitude smaller than the total thermal energy.

\subsection{Hybrid jets}

\begin{figure*}
	\centering
	\includegraphics[width=16 cm]{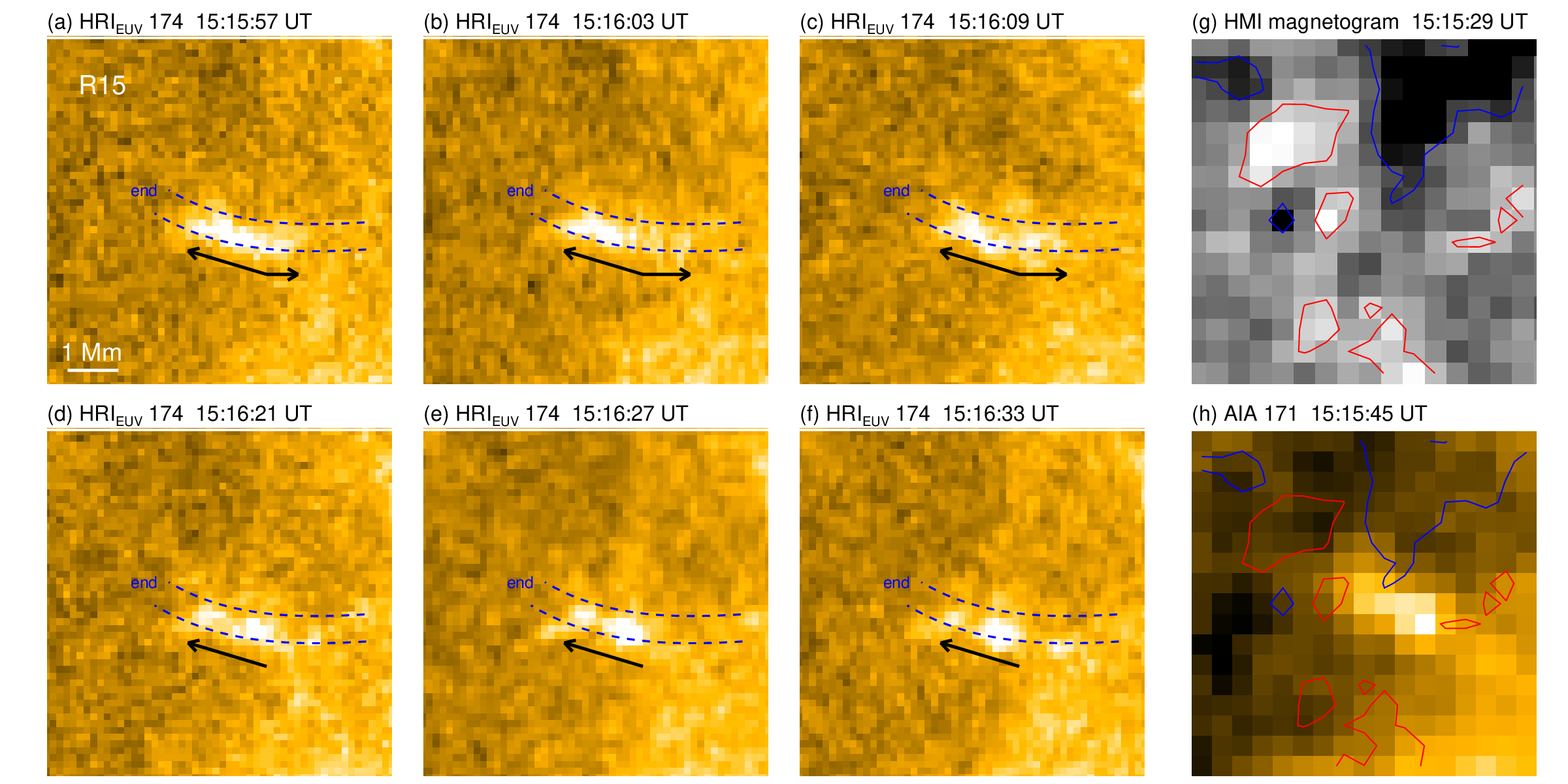}
	\caption{ (a)--(f) The temporal evolution of R15 jets. The black arrows indicate the propagation directions of jets. (g) HMI LOS magnetogram. The red and blue contours represent magnetic field strength with the levels of $\pm$10 G, respectively. (h) AIA 171 \AA\ map with overplotted HMI contours.}
	\label{figure5}%
\end{figure*}

\begin{figure*}
	\centering
	\includegraphics[width=16 cm]{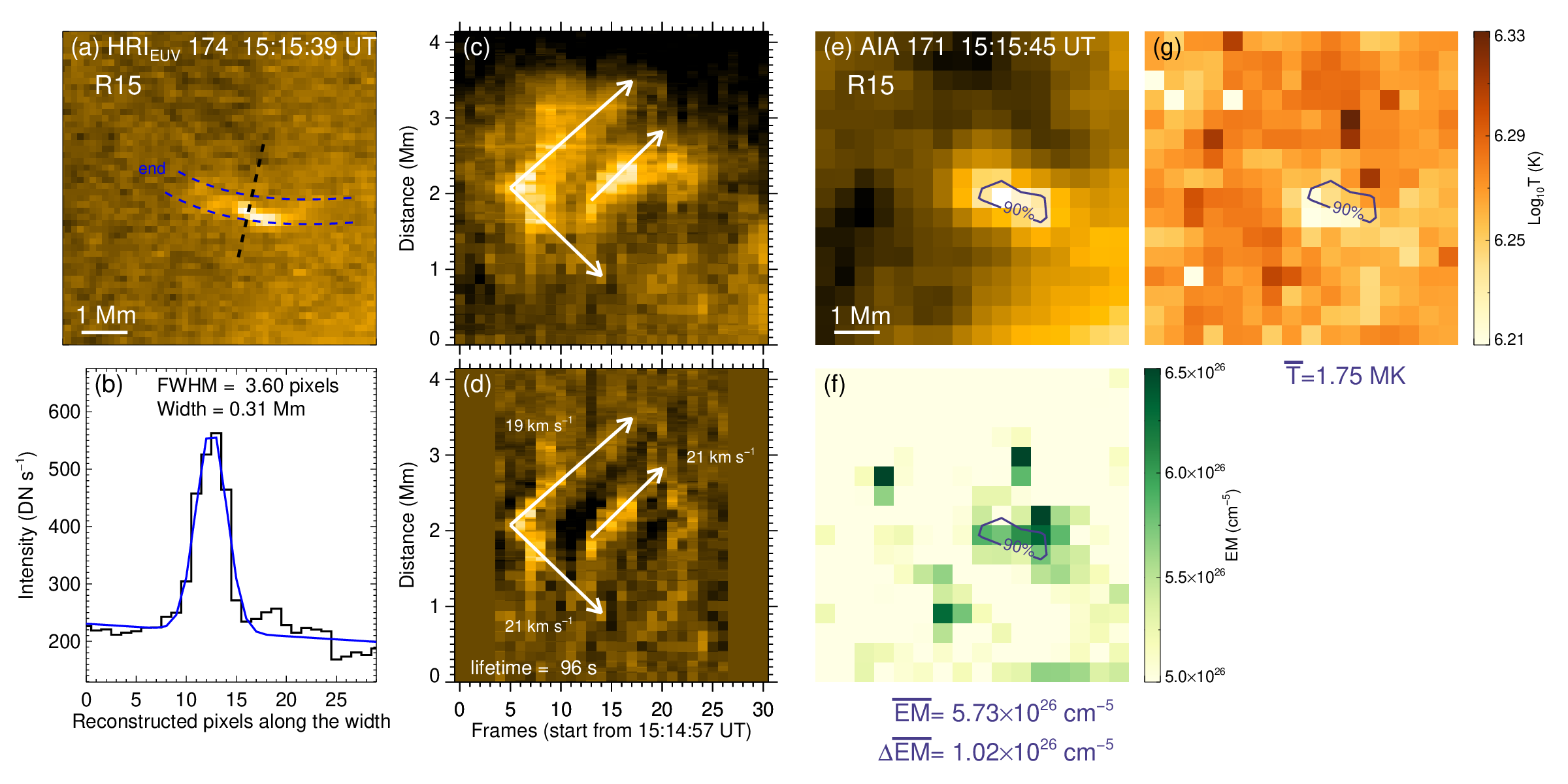}
	\caption{ (a)--(g) Similar to Fig.~\ref{figure3}, the results for R15 jets.}
	\label{figure6}%
\end{figure*}

Besides the unidirectional and bidirectional jets, the third type of jets was also detected. In this study, we refer to it as hybrid jets, which are mixtures of former two types. Fig.~\ref{figure5} and~\ref{figure6} give an example of the third type that is taken in R15. Fig.~\ref{figure5}(a)--(f) present the temporal evolution of jets. The first eruption is bidirectional, followed by a unidirectional propagation, as indicated by the black arrows. Fig.~\ref{figure5}(g) gives the HMI LOS magnetogram, and the red and blue contours represent magnetic field strength with the levels of $\pm$10 G, respectively. Fig.~\ref{figure5}(h) shows AIA 171 \AA\ map with overplotted HMI contours. From this we can see that the jet-base is locates in a weak opposite-polarity region, which is surrounded by several patches with magnetic field strength greater than 10 G. The magnetic flux cancellation in opposite-polarity regions is usually the trigger of quiet-region coronal jet eruptions \citep{Panesar2016ApJ...832L...7P}. The phenomenon of hybrid jets may be caused by the changes of magnetic field configuration during the first reconnection, which result in the subsequent unidirectional jet.

The total lifetime of hybrid jets is defined as the time interval between the initiation of first blob and the termination of last blob, it is 96 s in this event. The jet length is around 2 Mm based on Fig.~\ref{figure6}(a). Using the Equation \ref{ne}, \ref{thermal}, and \ref{thermalflux}, the electron density, thermal energy flux, and total thermal energy, are $1.81 \times 10^{9}\ \mathrm{cm^{-3}}$, $1.06 \times 10^{5}\ \mathrm{erg\ cm^{-2}\ s^{-1}}$, and $1.98 \times 10^{23}\ \mathrm{erg}$, respectively. The thermal energy flux accounts for $1/3$ of the energy required to heat the quiet-Sun corona. The total thermal energy is also in the range of $10^{23}$--$10^{24}\ \mathrm{erg}$.
 
 The respective lifetime of three blobs is 78, 60, and 48 s, with speed of $\sim$$19$, $\sim$$21$, and $\sim$$21\ \mathrm{km\ s^{-1}}$. Using the Equation \ref{kineticflux} and \ref{totalkinetic}, the kinetic energy fluxes are $0.10 \times 10^{5}$, $0.14 \times 10^{5}$, and $0.14 \times 10^{5}\ \mathrm{erg\ cm^{-2}\ s^{-1}}$, as well as the total kinetic energy is $1.76 \times 10^{21}\ \mathrm{erg}$. 
 
 Using the same method, the jets in other 19 regions are analyzed and shown in Appendix~\ref{appendix}. In order to display the jet contour appropriately, the contour level is chosen individually for every jet within the intensity range of 50--90\%. Their parameters and energies are listed in Table~\ref{table1}.

\begin{table*}
	\caption{Parameters and the calculated energies of the small-scale jets.}             
	\label{table1}      
	\centering          
	\begin{tabular}{l c c c c c c c c c c}     
		\hline\hline       
		Label & $w$\tablefootmark{a}  & $T$\tablefootmark{b} & $\tau_\mathrm{{tot}}$\tablefootmark{c}($\tau$\tablefootmark{d})  & $v$\tablefootmark{e} &$n_\mathrm{e}$\tablefootmark{f}&$F_\mathrm{k}$\tablefootmark{g}& $E_\mathrm{k}$\tablefootmark{h}& $F_\mathrm{{th}}$\tablefootmark{i}& $E_\mathrm{{th}}$\tablefootmark{j}  & Type\tablefootmark{k}\\ 
		                  &(Mm) & (MK) & (s) & ($\mathrm{km\ s^{-1}}$) & ($\mathrm{cm^{-3}}$) &($ \mathrm{erg\ cm^{-2}\ s^{-1}} $)& (erg)
		                  &($ \mathrm{erg\ cm^{-2}\ s^{-1}} $)& (erg) & \\ 
		
		\hline                    
		R1 & 0.39 & 1.63 & 36 & 120 & 1.62 $\times\ 10^{9}$ & 23.33 $\times\ 10^{5}$ & 1.00 $\times\ 10^{23}$ & 2.96 $\times\ 10^{5}$ & 3.91 $\times\ 10^{23}$ & $\mathit{1}$\\
		
		R2 & 0.31 & 1.39    & 108 &  & 2.04 $\times\ 10^{9}$ & (1.55--12.84) $\times\ 10^{5}$ & 6.08 $\times\ 10^{22}$ & 0.84 $\times\ 10^{5}$ & 2.22 $\times\ 10^{23}$ & $\mathit{2}$\\
						 &  &    & (30, 30, & 48, 45,  &  &  &   & & &\\
						 &  &    & 24, 24, &  91, 69, &  &  &   & & &\\
					     &  &    & 24, 24) & 72, 64 &  &  &   & & &\\
		
	    R3 & 0.34 & 1.67 & 42 & 64 & 1.16 $\times\ 10^{9}$ & 2.55 $\times\ 10^{5}$ & 9.71 $\times\ 10^{21}$ & 1.63 $\times\ 10^{5}$ & 1.46 $\times\ 10^{23}$ & $\mathit{1}$\\
	    
		R4\_a & 0.43 & 1.61   & 66 &   & 2.19 $\times\ 10^{9}$  & 0.32 $\times\ 10^{5}$ & 5.60 $\times\ 10^{21}$ & 2.38 $\times\ 10^{5}$ &  5.30 $\times\ 10^{23}$ & $\mathit{2}$\\
		&  &    & (66, 54) & 26, 26 &  &  &   &  & &\\
		
		R4\_b  & 0.29 & 1.73   & 30 &    & 1.41 $\times\ 10^{9}$ & (2.68--4.79) $\times\ 10^{5}$ & 9.94 $\times\ 10^{21}$ & 2.45 $\times\ 10^{5}$ &  1.00 $\times\ 10^{23}$ & $\mathit{1}$\\
		&  &    & (18, 24) & 74, 61 &  &  &   & & &\\
		
		R5 & 0.43 & 1.58   & 144 &   & 4.97 $\times\ 10^{9}$ & (77.2--189.7) $\times\ 10^{5}$ & 3.83 $\times\ 10^{24}$ & 2.42 $\times\ 10^{5}$ & 4.24 $\times\ 10^{24}$ & $\mathit{1}$\\
		&  &    & (48, 66, & 164, 166,  &  &  &   & & &\\
		&  &    & 36, 24) &  129, 123 &  &  &   & & &\\
		
		R6 & 0.36 & 1.73 & 48 & 65 & 1.14 $\times\ 10^{9}$ & 2.62 $\times\ 10^{5}$ & 1.28 $\times\ 10^{22}$ & 1.53 $\times\ 10^{5}$ & 1.67 $\times\ 10^{23}$ & $\mathit{1}$\\
		
		R7 & 0.57 & 1.65   & 90 &   & 0.98 $\times\ 10^{9}$ & (19.3--31.7) $\times\ 10^{5}$ & 8.24 $\times\ 10^{23}$ & 1.06 $\times\ 10^{5}$ & 1.03 $\times\ 10^{24}$  & $\mathit{1}$\\
		&  &    & (42, 48, & 157, 145,  &  &  &   & & &\\
		&  &    & 36) &  133 &  &  &   & & &\\
		
		R8 & 0.49 & 1.70   & 108 &   & 1.55 $\times\ 10^{9}$ & (4.64--26.0) $\times\ 10^{5}$ & 5.62 $\times\ 10^{23}$ & 1.24 $\times\ 10^{5}$ & 1.65 $\times\ 10^{24}$ & $\mathit{3}$\\
		&  &    & (42, 54, & 126, 106,  &  &  &   & & &\\
		&  &    & 48, 24) &  115, 71 &  &  &   & & &\\
		
		R9 & 0.34 & 1.67   & 36 &    & 1.58 $\times\ 10^{9}$ & (0.57--0.62) $\times\ 10^{5}$ & 2.94 $\times\ 10^{21}$ & 2.58 $\times\ 10^{5}$ &  2.48 $\times\ 10^{23}$ & $\mathit{1}$\\
		&  &    & (18, 36) & 35, 36 &  &  &   & & &\\
		
		R10 & 0.35 & 1.61   & 72 &    & 2.39 $\times\ 10^{9}$ & (25.1--32.0) $\times\ 10^{5}$ & 1.87 $\times\ 10^{23}$ & 1.94 $\times\ 10^{5}$ &  6.13 $\times\ 10^{23}$ & $\mathit{1}$\\
		&  &    & (24, 42) & 108, 117 &  &  &   & & &\\
		
		R11 & 0.32 & 1.74 & 78 &   & 1.19 $\times\ 10^{9}$ & (0.03--0.12) $\times\ 10^{5}$ & 6.22 $\times\ 10^{20}$ & 0.88 $\times\ 10^{5}$ & 1.72 $\times\ 10^{23}$ & $\mathit{3}$\\
		&  &    & (30, 54, & 23, 18,  &  &  &   & & &\\
		&  &    & 30) &  15 &  &  &   & & &\\
		
		R12 & 0.32 & 1.78    & 30  & 98 & 1.06 $\times\ 10^{9}$  & 8.34 $\times\ 10^{5}$ & 2.01 $\times\ 10^{22}$ & 2.08 $\times\ 10^{5}$ &  9.43 $\times\ 10^{22}$ & $\mathit{1}$\\
		
		R13 & 0.34 & 1.75   & 30 &    & 1.25 $\times\ 10^{9}$ & (0.57--5.34) $\times\ 10^{5}$ & 1.61 $\times\ 10^{22}$ & 2.56 $\times\ 10^{5}$ &  2.46 $\times\ 10^{23}$ & $\mathit{2}$\\
						 &  &    & (30, 30) & 80, 38 &  &  &   & & &\\
					    
		R14  & 0.26 & 1.76   & 66 &   & 4.67 $\times\ 10^{9}$ & (0.47--1.08) $\times\ 10^{5}$ & 6.10 $\times\ 10^{21}$ & 1.34 $\times\ 10^{5}$ & 1.81 $\times\ 10^{23}$ & $\mathit{2}$\\
		&  &    & (42, 42, & 34, 31,  &  &  &   & & &\\
		&  &    & 36, 36) &  41, 38 &  &  &   & & &\\
		
		R15 & 0.31 & 1.75    & 96 &   & 1.81 $\times\ 10^{9}$ & (0.10--0.14) $\times\ 10^{5}$ & 1.76 $\times\ 10^{21}$ & 1.06 $\times\ 10^{5}$ & 1.98 $\times\ 10^{23}$ & $\mathit{3}$\\
				 &  &    & (78, 60, & 19, 21, &  &  &   & & &\\
				 &  &    & 48) & 21 &  &  &   & & &\\
		
		R16 & 0.33 & 1.63   & 60 &    & 2.22 $\times\ 10^{9}$ & (0.41--1.02) $\times\ 10^{5}$ & 6.79 $\times\ 10^{21}$ & 2.06 $\times\ 10^{5}$ &  4.49 $\times\ 10^{23}$ & $\mathit{2}$\\
		&  &    & (54, 60) & 38, 28 &  &  &   & & &\\
		
		R17 & 0.48 & 1.68    & 120 &  & 1.63 $\times\ 10^{9}$ & (0.24--23.5) $\times\ 10^{5}$ & 6.03 $\times\ 10^{23}$ & 1.13 $\times\ 10^{5}$ & 1.43 $\times\ 10^{24}$ & $\mathit{2}$\\
		&  &    & (42, 36, & 116, 62,  &  &  &   & & &\\
		&  &    & 66, 84, &  120, 26, &  &  &   & & &\\
		&  &    & 24, 24, & 110, 58 &  &  &   & & &\\
		&  &    & 30, 30) & 82, 43 &  &  &   & & &\\
		
		R18 & 0.36 & 1.65   & 66 &   & 1.50 $\times\ 10^{9}$ & (0.13--3.44) $\times\ 10^{5}$ & 1.58 $\times\ 10^{22}$ & 1.40 $\times\ 10^{5}$ & 2.61 $\times\ 10^{23}$ & $\mathit{2}$\\
		&  &    & (24, 30, & 65, 45,  &  &  &   & & &\\
		&  &    & 24, 24) &  49, 22 &  &  &   & & &\\
		
		R19 & 0.33 & 1.77   & 42 &   & 1.54 $\times\ 10^{9}$ & (0.46--6.82) $\times\ 10^{5}$ & 3.23 $\times\ 10^{22}$ & 2.21 $\times\ 10^{5}$ & 2.41 $\times\ 10^{23}$ & $\mathit{2}$\\
		&  &    & (24, 24, & 81, 61,  &  &  &   & & &\\
		&  &    & 24, 30) &  75, 33 &  &  &   & & &\\
		
		R20 & 0.38 & 1.90   & 102 &    & 2.61 $\times\ 10^{9}$ & (32.3--69.3) $\times\ 10^{5}$ & 4.84 $\times\ 10^{23}$ & 1.91 $\times\ 10^{5}$ &  1.16 $\times\ 10^{24}$ & $\mathit{1}$\\
		&  &    & (42, 42) & 147, 114 &  &  &   & & & \\
		
		R21 & 0.40 & 1.78    & 54  & 37  & 1.62 $\times\ 10^{9}$ & 0.69 $\times\ 10^{5}$ & 4.65 $\times\ 10^{21}$ & 2.21 $\times\ 10^{5}$ &  3.75 $\times\ 10^{23}$ & $\mathit{1}$\\
		
		R22 & 0.27 & 1.70   & 156 &   & 2.42 $\times\ 10^{9}$ & (0.55--9.96) $\times\ 10^{5}$ & 6.15 $\times\ 10^{22}$ & 0.74 $\times\ 10^{5}$ & 5.85 $\times\ 10^{24}$ & $\mathit{3}$\\
		&  &    & (36, 30, & 75, 58,  &  &  &   & & &\\
		&  &    & 18, 54 &  30, 79, &  &  &   & & &\\
		&  &    & 48) &  47 &  &  &   & & &\\

		\hline                  
	\end{tabular}
	\tablefoot{
	\tablefoottext{a}{Width.}
	\tablefoottext{b}{Mean EM-weighted temperature.}
	\tablefoottext{c}{The total lifetime ($\tau_\mathrm{{tot}}$) was uesd to calculate the thermal energy flux ($F_\mathrm{th}$).}
	\tablefoottext{d}{Each blob lifetime ($\tau$) was used to calculate the total kinetic energy ($E_\mathrm{k}$).}
	\tablefoottext{e}{Velocity of each blob.}
	\tablefoottext{f}{Mean electron density.}
	\tablefoottext{g}{Kinetic energy flux.}
	\tablefoottext{h}{Total kinetic energy.}
	\tablefoottext{i}{Thermal energy flux.}
	\tablefoottext{j}{Total thermal energy.}
	\tablefoottext{k}{Type1: Unidirectional, 2: Bidirectional, 3: Hybrid.}
	
}

\end{table*}

\section{Discussion and Conclusions}  \label{discuss}
\begin{figure}
	\centering
	\includegraphics[width=\hsize]{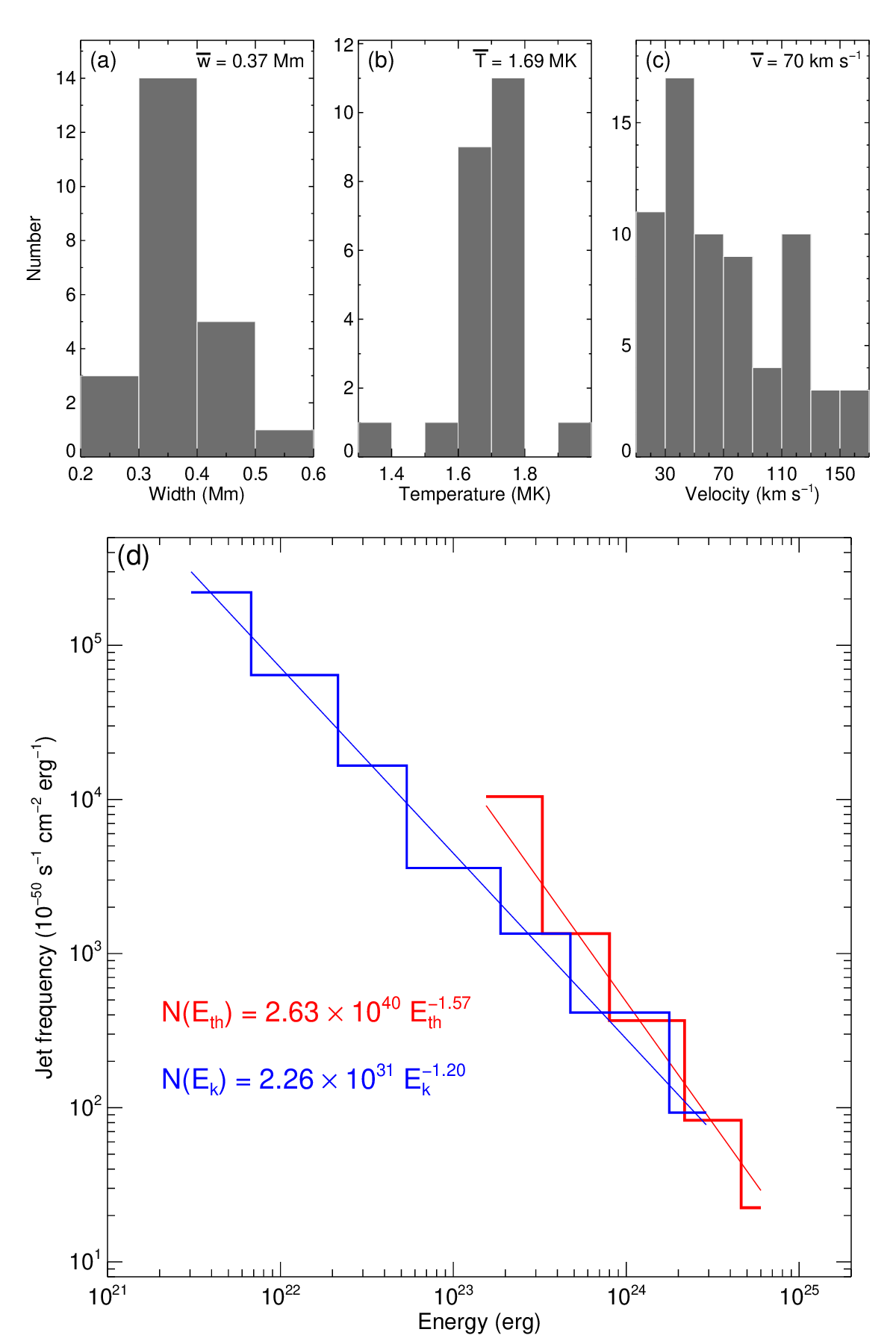}
	\caption{ (a)--(c) The histograms of width, temperature, and velocity, respectively. The average values are given. (d) Jet frequency distribution and the fitting results in a normalized scale in units of $10^{-50}$ jets per unit time ($\mathrm{s^{-1}}$), unit area ($\mathrm{cm^{-2}}$), and unit energy ($\mathrm{erg^{-1}}$). The power-law distributions $N(E) \propto$ $E^{-\alpha}$ (red for thermal energy, blue for kinetic energy) are also given.}
	\label{figure7}%
\end{figure}

Using the high spatiotemporal resolution observation of HRI$_\mathrm{{EUV}}$ $174$ \AA\ of SolO/EUI, 23 small-scale jets were identified in a quiet-Sun region that was observed at 15:00:15--16:07:09 UT on 2023 March 30. Three types can be classified based on their motion behaviors, i.e., unidirectional, bidirectional, and hybrid jets. The third type may be caused by the changes of magnetic field configuration during the first reconnection, which affects the subsequent eruption. Most of these jets also showed the repeated blobs, which could arise from the quasi-periodic reconnection that have been observed in some solar flares \citep{Shi2022Univ....8..104S, Shi2022RAA....22j5017S, Li2022FrASS...932099L}. Combining the simultaneous observation from SDO/AIA, their physical parameters and energies were measured. According to the Table~\ref{table1}, Fig.~\ref{figure7}(a)--(c) plot the histograms of width, temperature, and velocity, with average value of $0.37$ Mm, $1.69$ MK, and $70\ \mathrm{km\ s^{-1}}$, respectively. These jets typically have the width of $\sim$$0.3$ Mm, the temperature of $\sim$$1.7$ MK, with velocity in a wide range of $\sim$$20$--$170\ \mathrm{km\ s^{-1}}$.

Most of these jets have the energy of $10^{23}$--$10^{24}\ \mathrm{erg}$, which is marginally smaller than the energy of typical nanoflares ($\sim$$10^{24}\ \mathrm{erg}$, e.g., \citealt{Parker1988ApJ...330..474P, Chitta2021A&A...647A.159C}). We also found that the kinetic energy of small-scale jets usually one to two orders of magnitude lower than the thermal energy, except for several relative intense jets, such as R5, R7, and R8, which showed the ambiguous inverted Y-shaped structures similar to standard or blowouts jets \citep{Moore2010ApJ...720..757M, Sterling2015Natur.523..437S}. As shown in Table \ref{table1}, the thermal energy fluxes of 23 jets are estimated to be (0.74--2.96) $\times 10^{5}\ \mathrm{erg\ cm^{-2}\ s^{-1}}$, which is almost on the same order as the energy flow required to heat the quiet-Sun corona ($3 \times 10^{5}\ \mathrm{erg\ cm^{-2}\ s^{-1}}$, \citealt{Withbroe1977ARA&A..15..363W}), although the kinetic energy fluxes are changing over a wide range due to their strong dependence on velocity. Additionally, based on the current 23 samples, we did not find significant discrepancies in energetics between the different jet types.

Fig.~\ref{figure7}(d) plots the jet frequency distribution and the fitting results in a normalized scale in units of $10^{-50}$ jets per unit time ($\mathrm{s^{-1}}$), unit area ($\mathrm{cm^{-2}}$), and unit energy ($\mathrm{erg^{-1}}$). The thermal energy (red) and kinetic energy (blue) both follow the power-law distribution $N(E) \propto$ $E^{-\alpha}$, with slope of $-1.57$ and $-1.20$, respectively. Note that the distribution shown here comes from relatively small sample size of 23 jets, thus the distribution only shows a rough approximation. Our results are roughly consistent with the composite flare frequency distribution as shown by \cite{Aschwanden2000ApJ...535.1047A}, who integrated multiple studies over the energy domain of $10^{24}$--$10^{32}$ erg \citep{Crosby1993SoPh..143..275C, Shimizu1997ApJ...486.1045S, Krucker1998ApJ...501L.213K, Parnell2000ApJ...529..554P}.

As shown by \cite{Withbroe1977ARA&A..15..363W}, the coronal energy losses in quiet-Sun region consist of three components: conduction flux ($2 \times 10^{5}\ \mathrm{erg\ cm^{-2}\ s^{-1}}$), radiative flux ($10^{5}\ \mathrm{erg\ cm^{-2}\ s^{-1}}$), and solar wind flux ($\lesssim$$5 \times 10^{4}\ \mathrm{erg\ cm^{-2}\ s^{-1}}$). The area of quiet-Sun region investigated in this study is $136 \times 136\ \mathrm{Mm^{2}}$ as seen in Fig.~\ref{figure1}, and thus the total energy required to sustain the heating of whole region during 15:00:15--16:07:09 UT is around $2 \times 10^{29}\ \mathrm{erg}$. Summing the total energy of the 23 events in Table~\ref{table1}, we can obtain a value of about $3\times 10^{25}\ \mathrm{erg}$, which means that the 23 events only account for a very small portion of the total energy demand during this observation. Considering that the 23 events are only a part of all the jets, there are still substantial jets in the image sequence of HRI$_\mathrm{{EUV}}$ $174$ \AA\ (movie\_hri174.mp4) that are difficult to be recognized because of relative strong background emissions or transient properties, or several jets occurring in the same region (such as R4\_a and R4\_b), but are not listed. It is reasonable to infer that those jets have the similar thermal energy fluxes as calculated here (comparable to $3 \times 10^{5}\ \mathrm{erg\ cm^{-2}\ s^{-1}}$), contributing the heating of local corona. Thus, actually the contribution of small-scale jets for coronal heating will become greater. Besides the energy input into corona, the dissipation mechanisms of energy is also important. The relevant dissipation mechanisms need to be investigated in future studies. In our observations, although these jets cannot provide sufficient energy to heat the whole quiet-Sun coronal region, they are likely to account for a significant portion of the energy demand in the local regions where the jets occur.

\begin{acknowledgements}
We thank the referee for valuable comments. This work is supported by the Strategic Priority Research Program of the Chinese Academy of Sciences, Grant No. XDB0560000. This work is also funded by the National Key R\&D Program of China2022YFF0503002 (2022YFF0503000), NSFC under grants 12073081. Solar Orbiter is a space mission of international collaboration between ESA and NASA, operated by ESA. We appreciate the teams of SDO for their open data-use policy.
\end{acknowledgements}

%
\bibliographystyle{aa} 
\bibliography{myreferences} 
%

\begin{appendix} 
	\section{The remaining small-scale jets} \label{appendix}

\begin{figure*}[!b]
	\centering
		\subfigure{
		\includegraphics[width=14 cm]{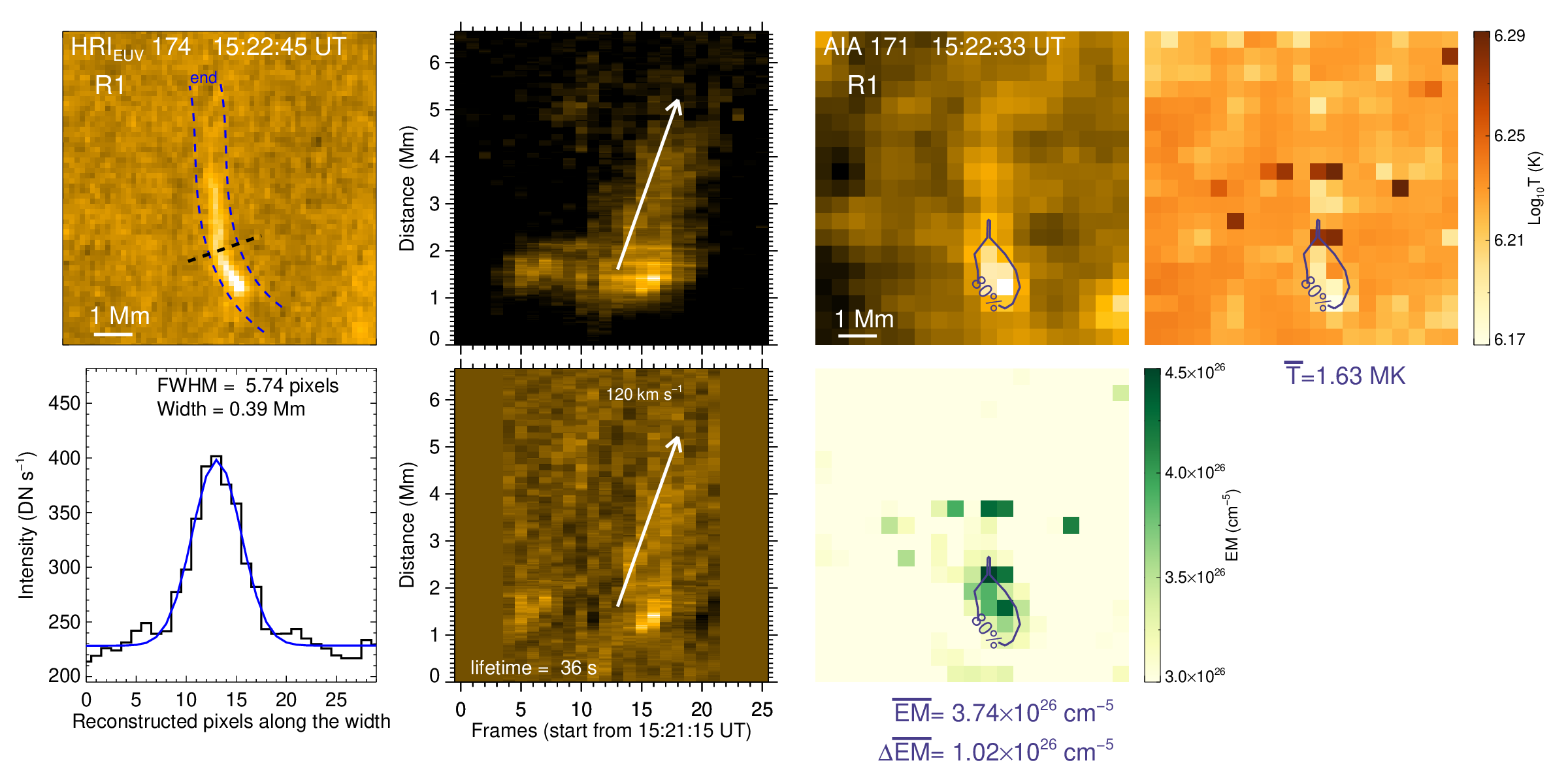}
		}
		\subfigure{
		\includegraphics[width=14 cm]{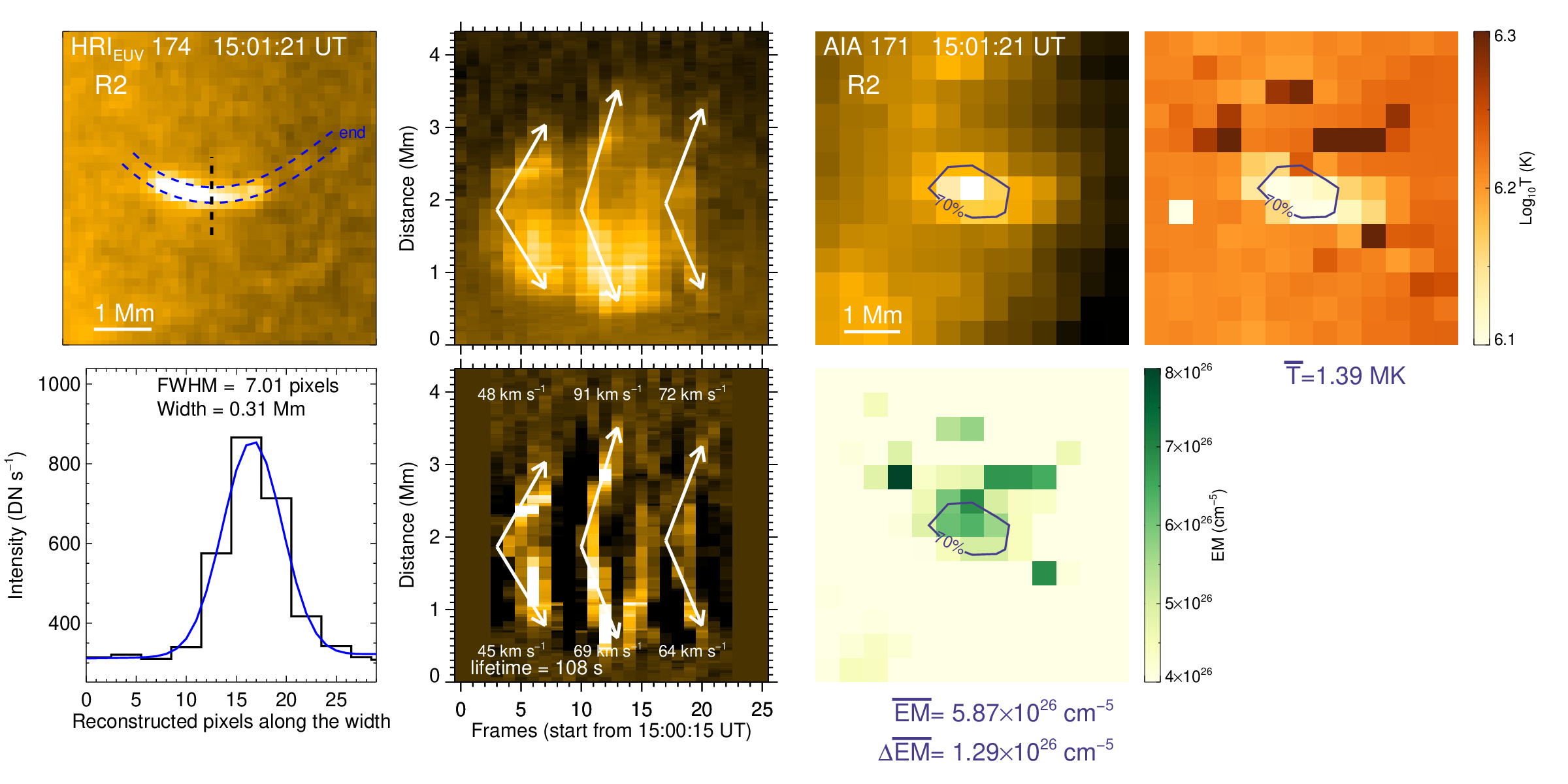}
	    }
		\subfigure{
		\includegraphics[width=14 cm]{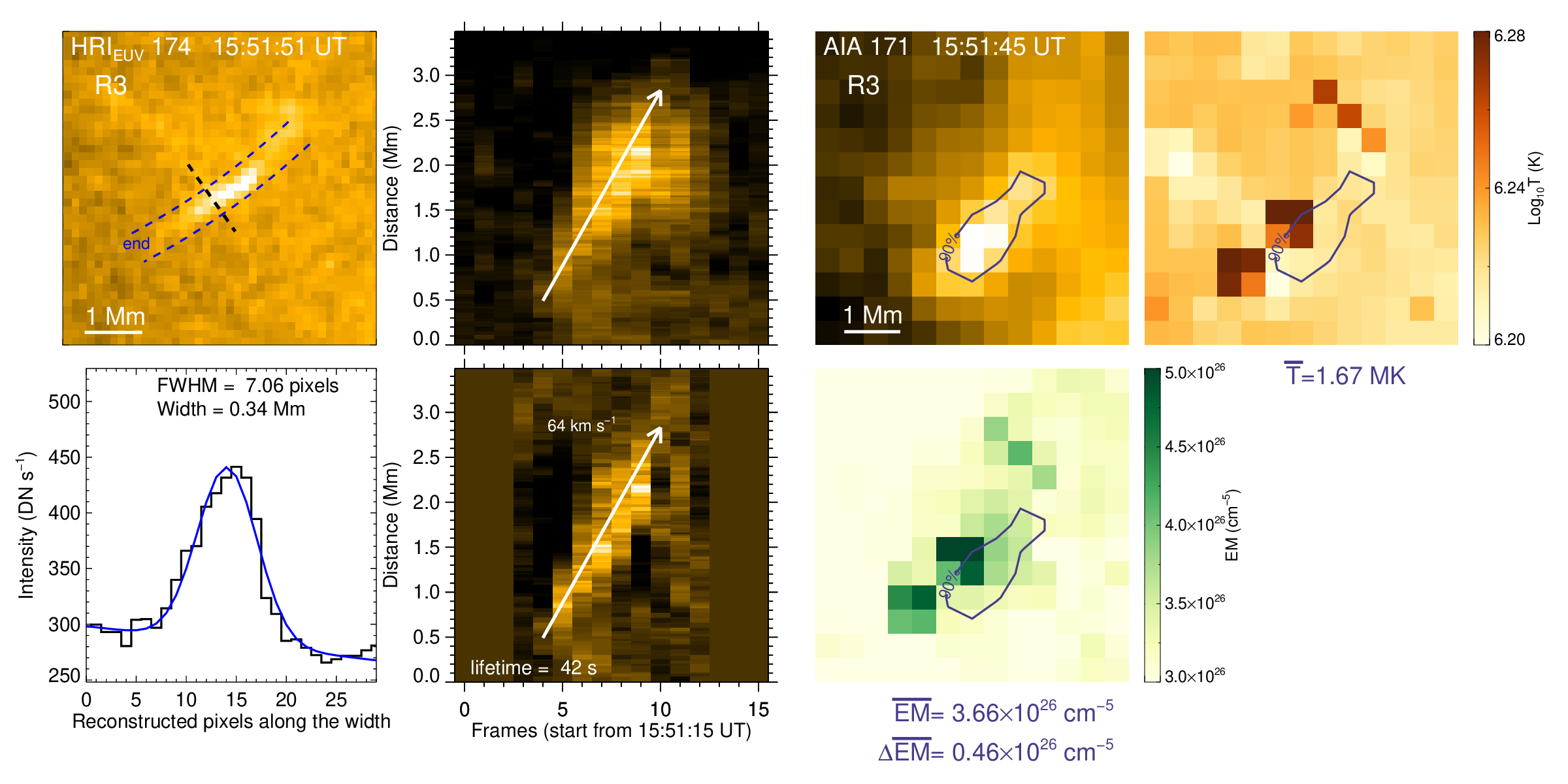}
		}
	\caption{Similar to Figure~\ref{figure3}, the results for R1, R2, and R3.}\label{figure_appenix1}
	\end{figure*}

\begin{figure*}[!b]
	\centering
	\subfigure{
		\includegraphics[width=15 cm]{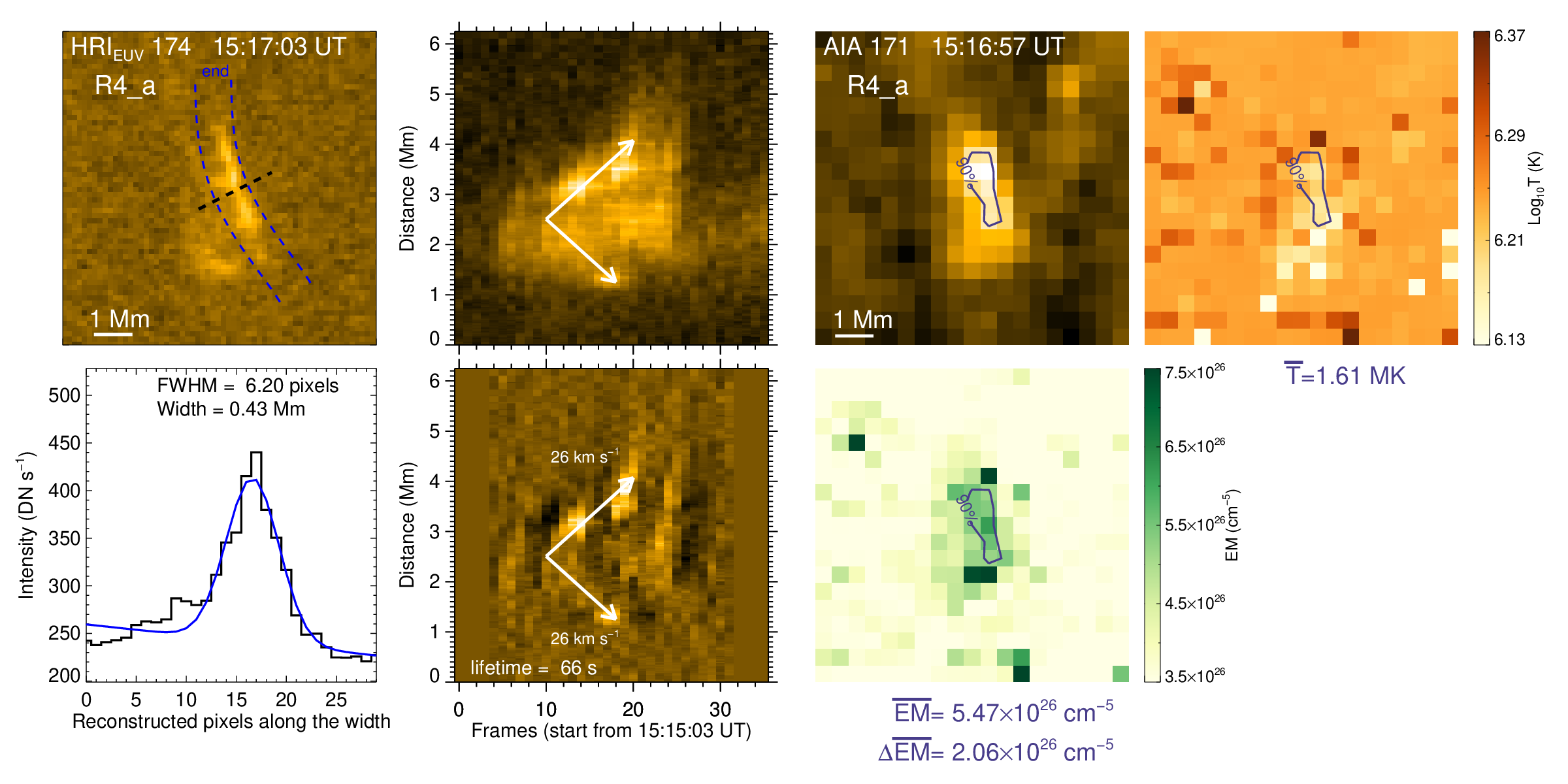}
	}
	\subfigure{
		\includegraphics[width=15 cm]{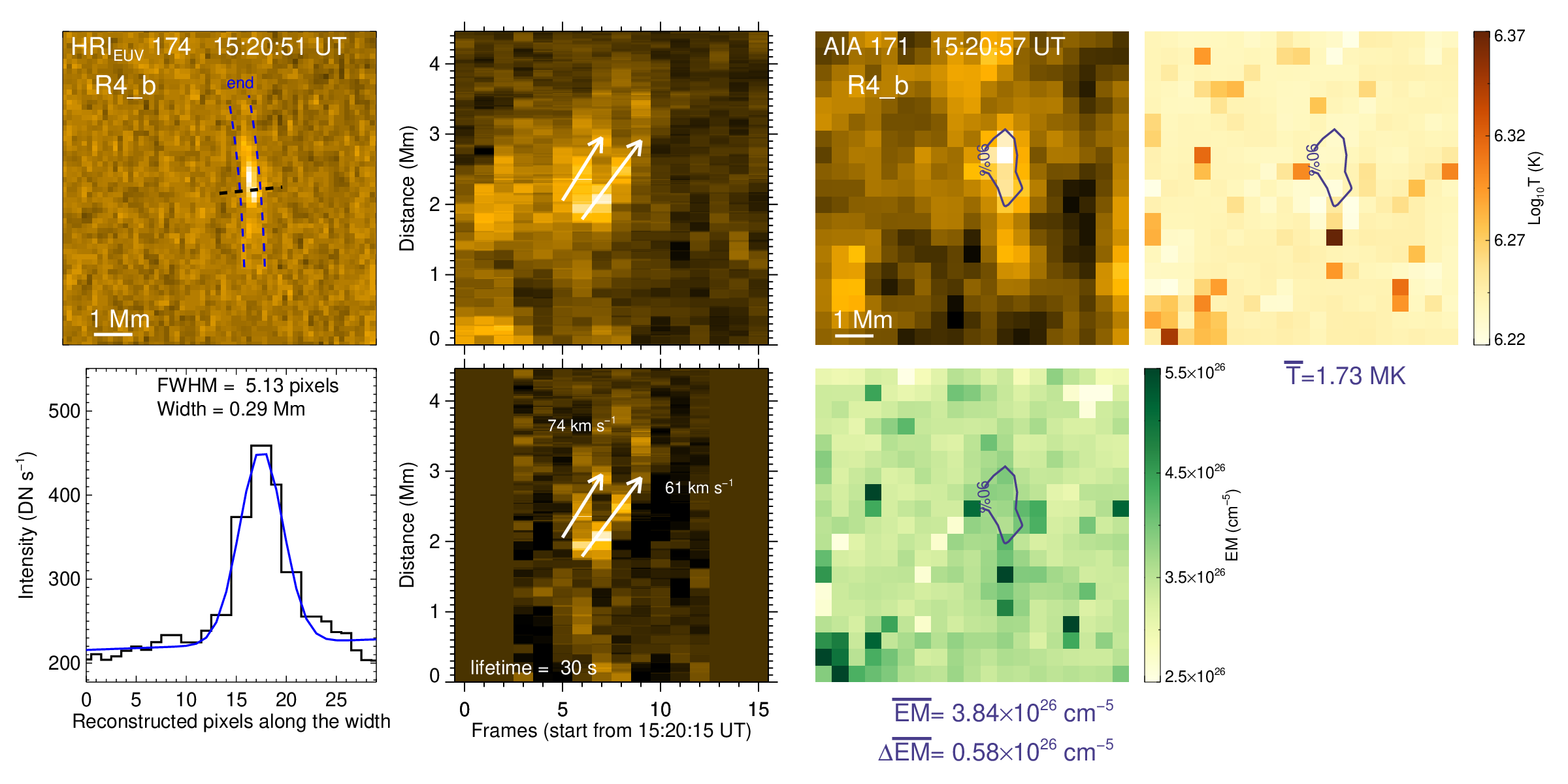}
	}
	\subfigure{
		\includegraphics[width=15 cm]{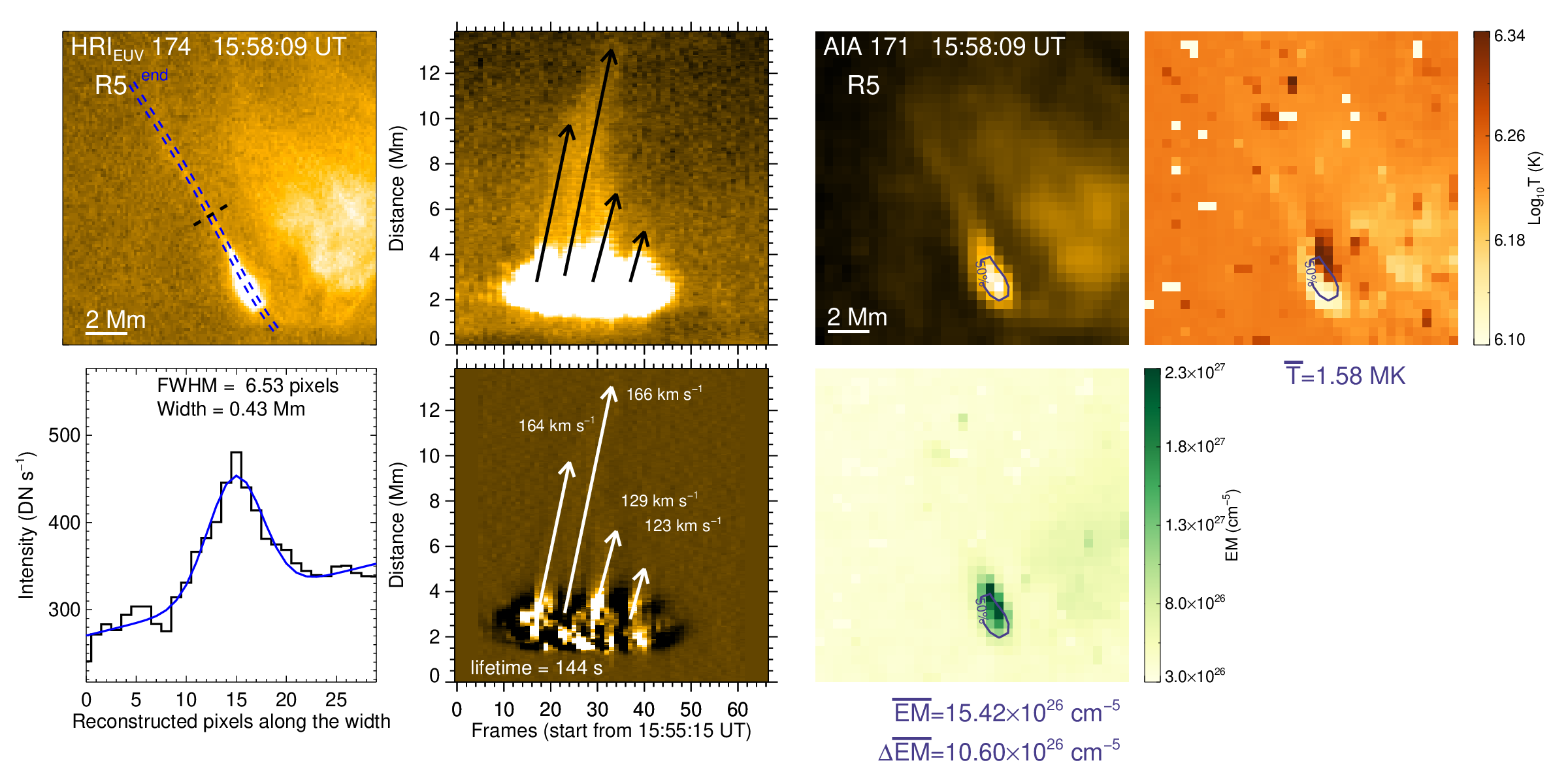}
	}
	\caption{Similar to Figure~\ref{figure3}, the results for R4\_a, R4\_b, and R5.}\label{figure_appenix2}
\end{figure*}

\begin{figure*}[!b]
	\centering
	\subfigure{
		\includegraphics[width=15 cm]{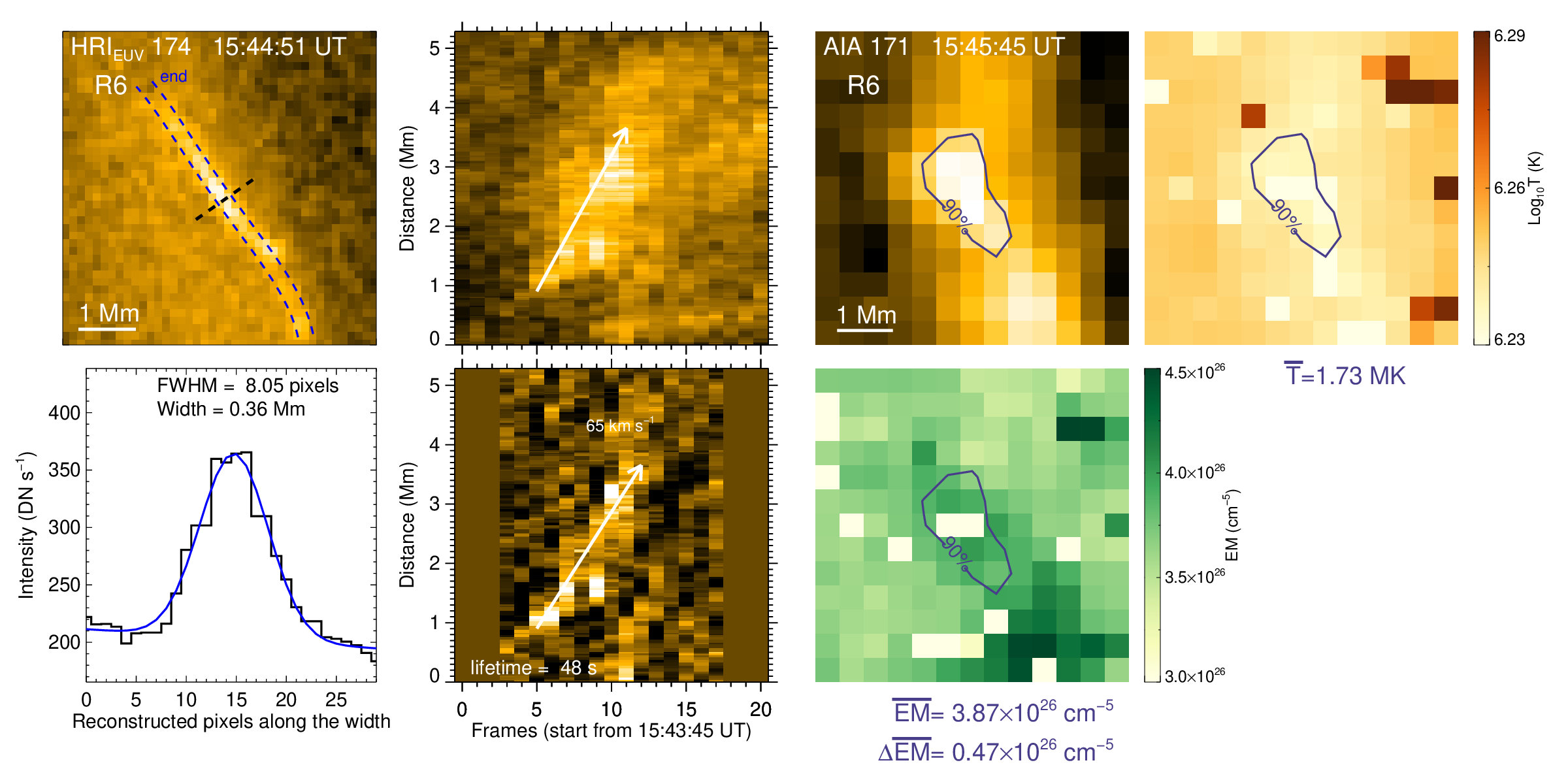}
	}
	\subfigure{
		\includegraphics[width=15 cm]{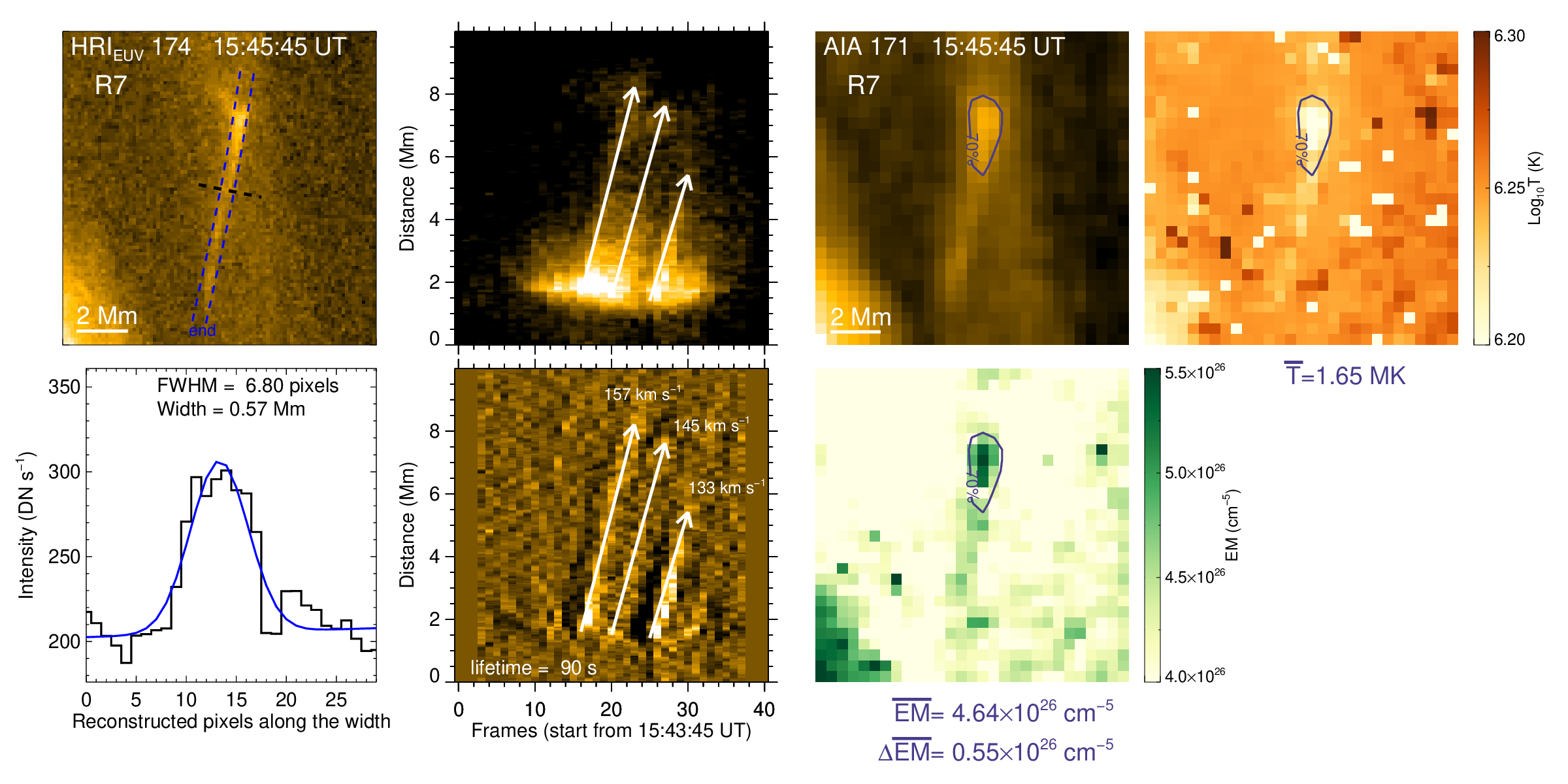}
	}
	\subfigure{
		\includegraphics[width=15 cm]{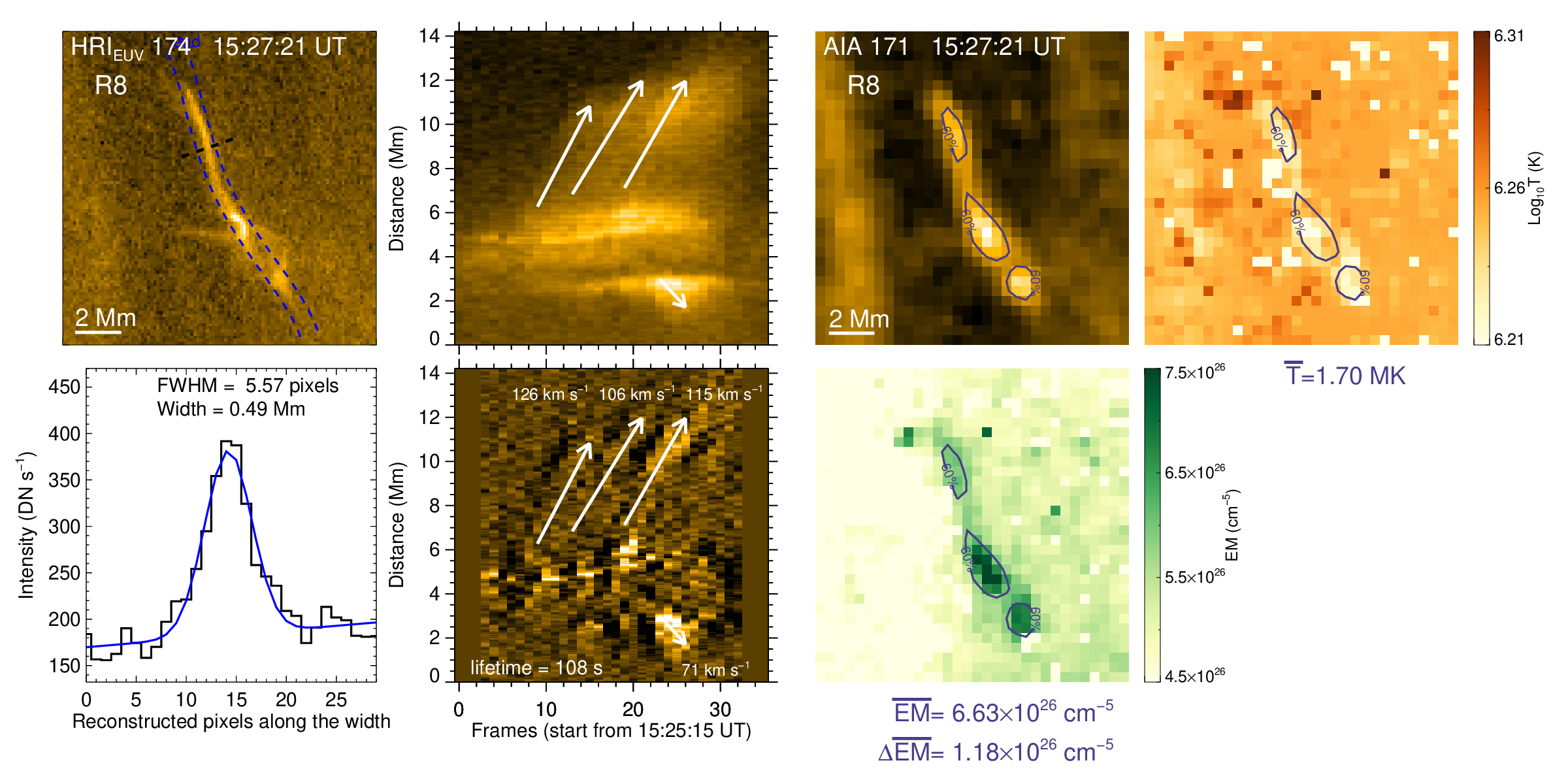}
	}
	\caption{Similar to Figure~\ref{figure3}, the results for R6, R7, and R8.}\label{figure_appenix3}
\end{figure*}

\begin{figure*}[!b]
	\centering
	\subfigure{
		\includegraphics[width=15 cm]{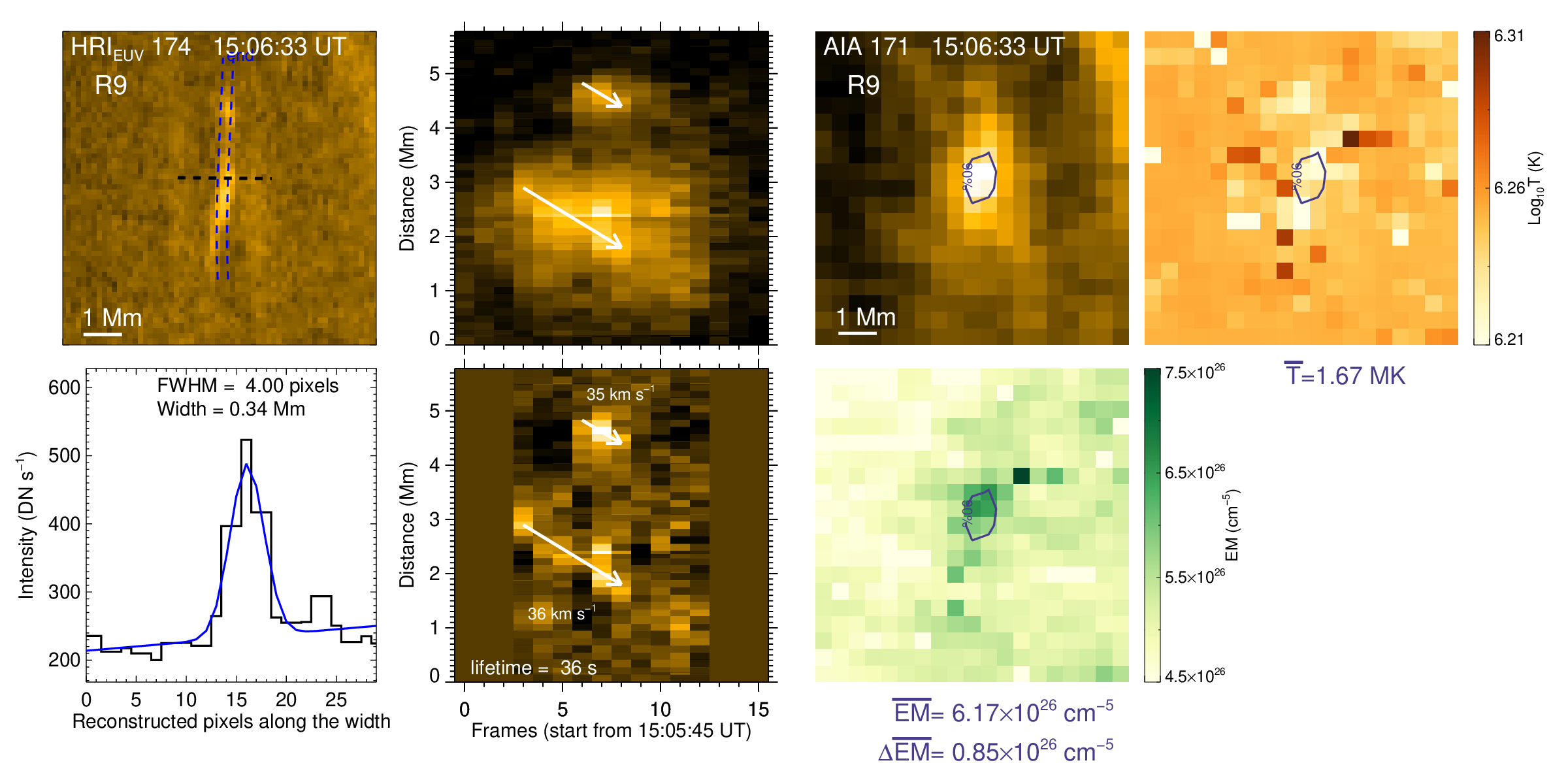}
	}
	\subfigure{
		\includegraphics[width=15 cm]{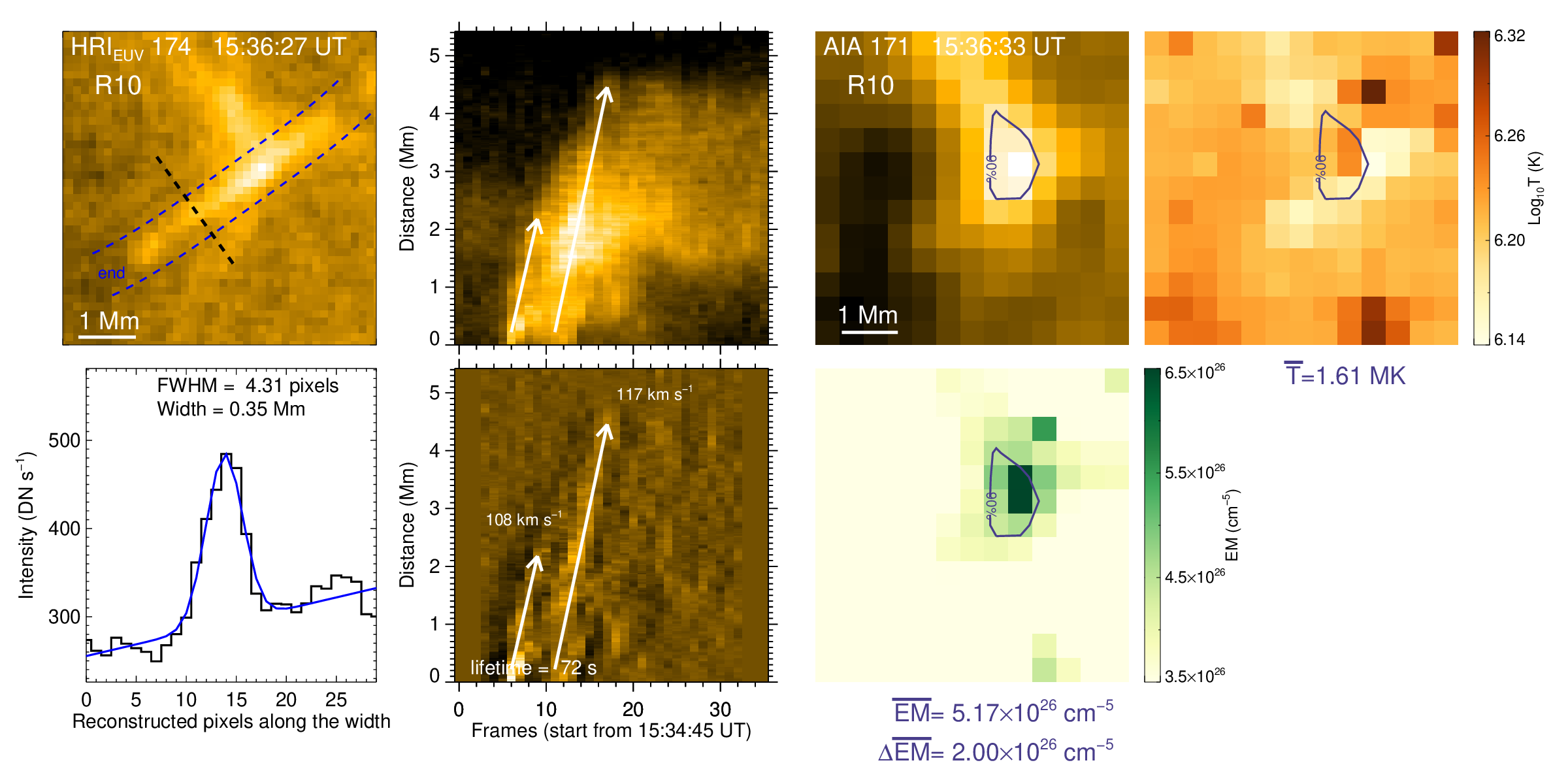}
	}
	\subfigure{
		\includegraphics[width=15 cm]{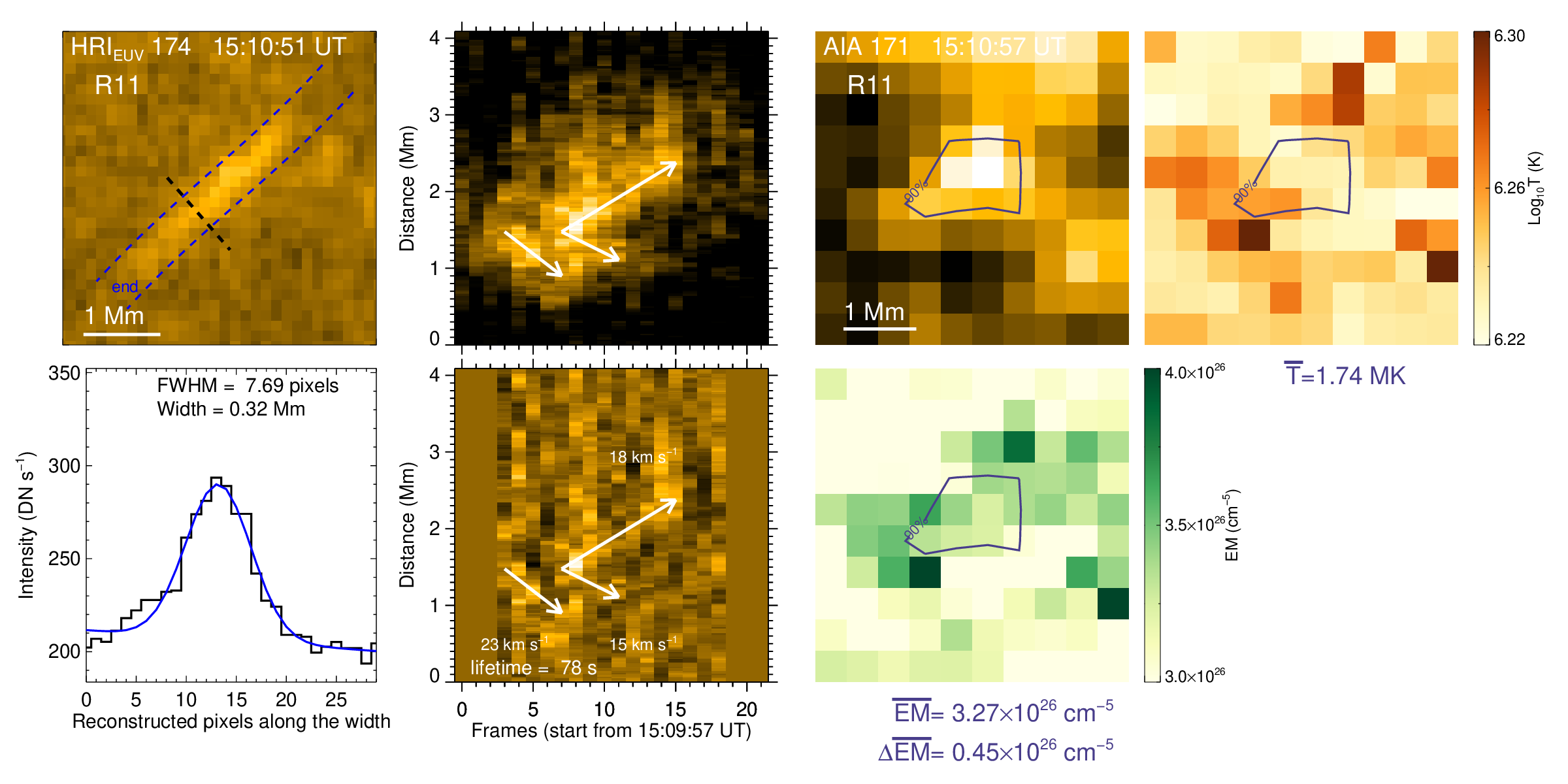}
	}
	\caption{Similar to Figure~\ref{figure3}, the results for R9, R10, and R11.}\label{figure_appenix4}
\end{figure*}

\begin{figure*}[!b]
	\centering
	\subfigure{
		\includegraphics[width=15 cm]{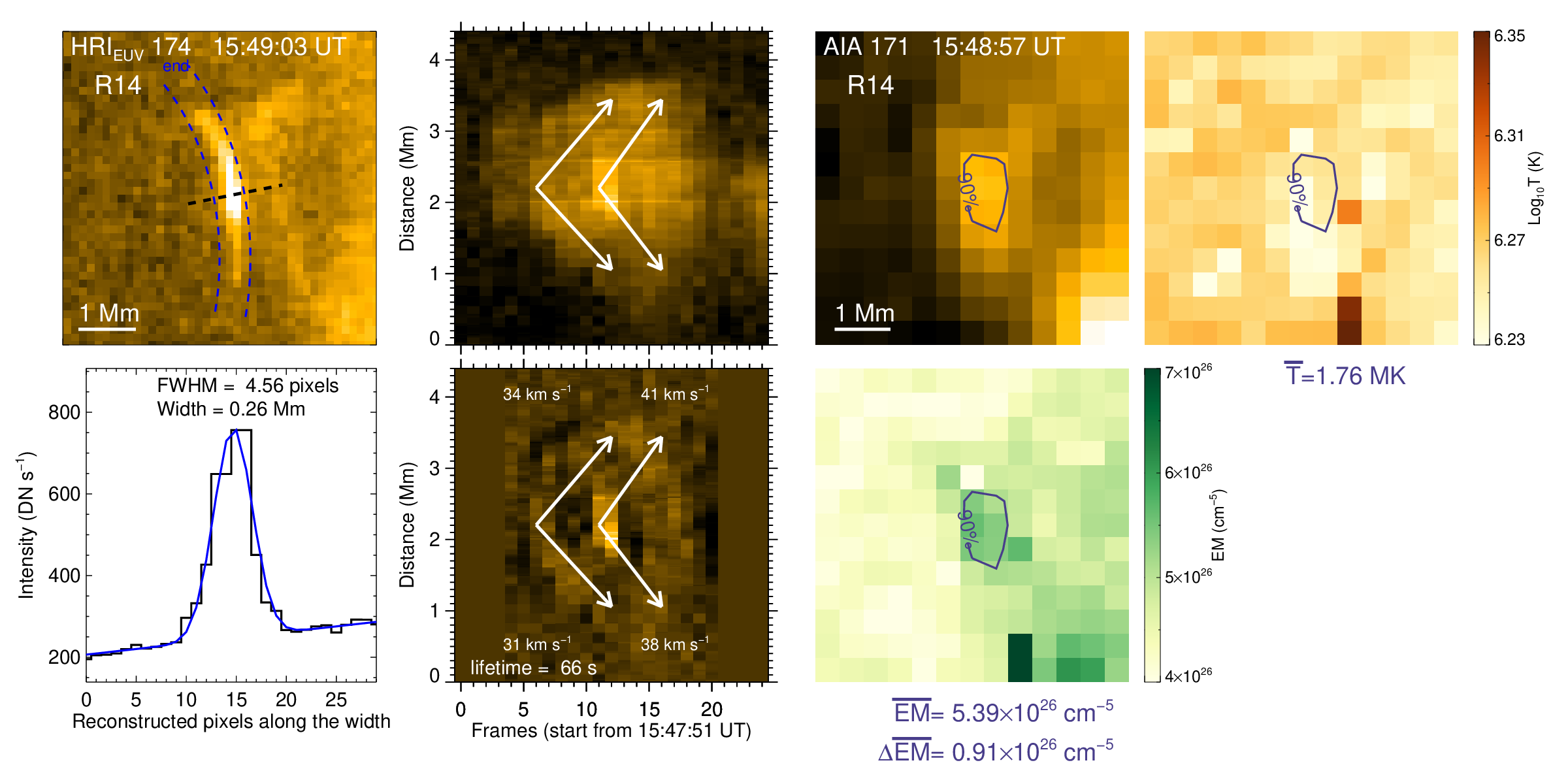}
	}
	\subfigure{
		\includegraphics[width=15 cm]{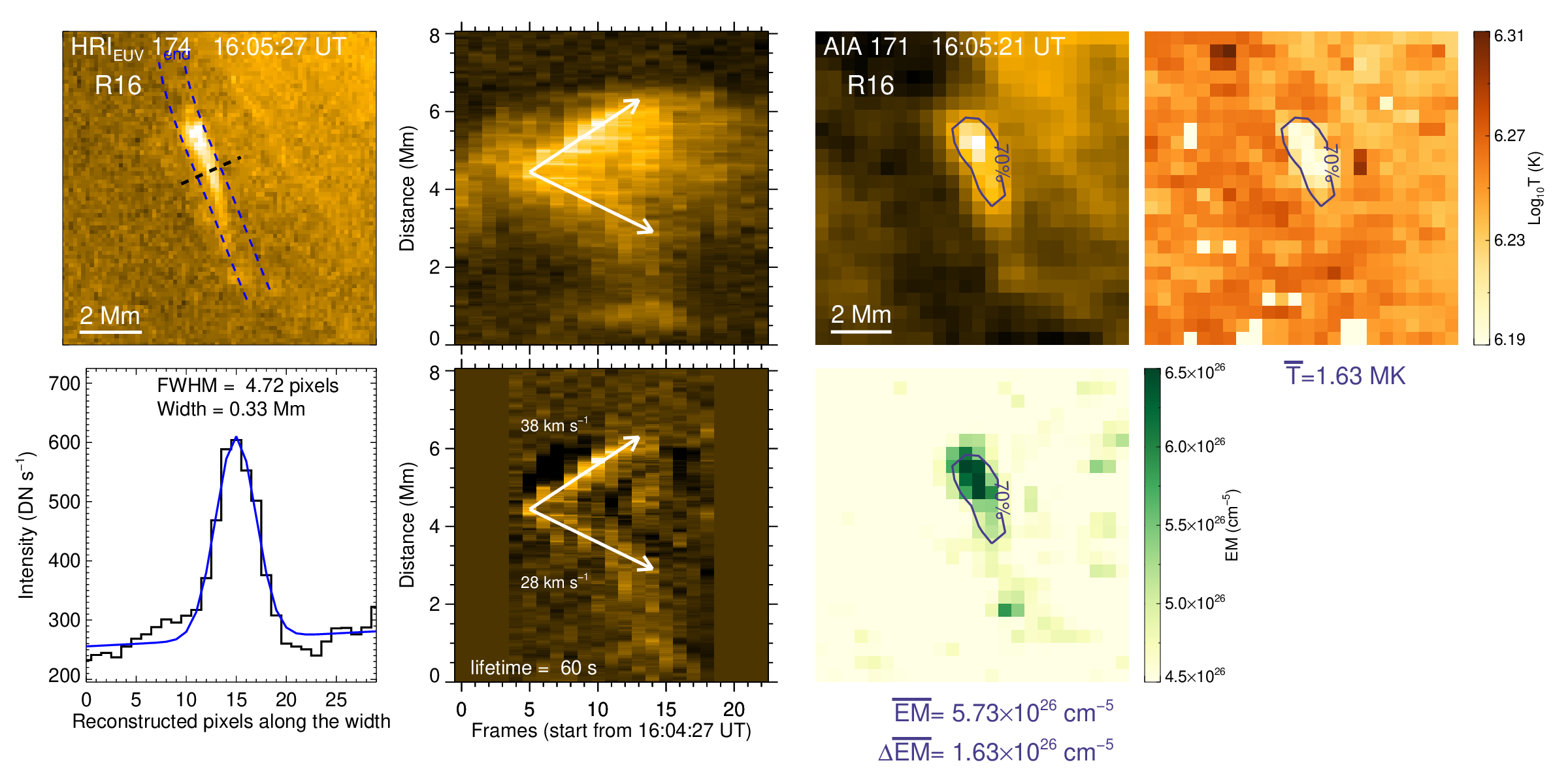}
	}
	\subfigure{
		\includegraphics[width=15 cm]{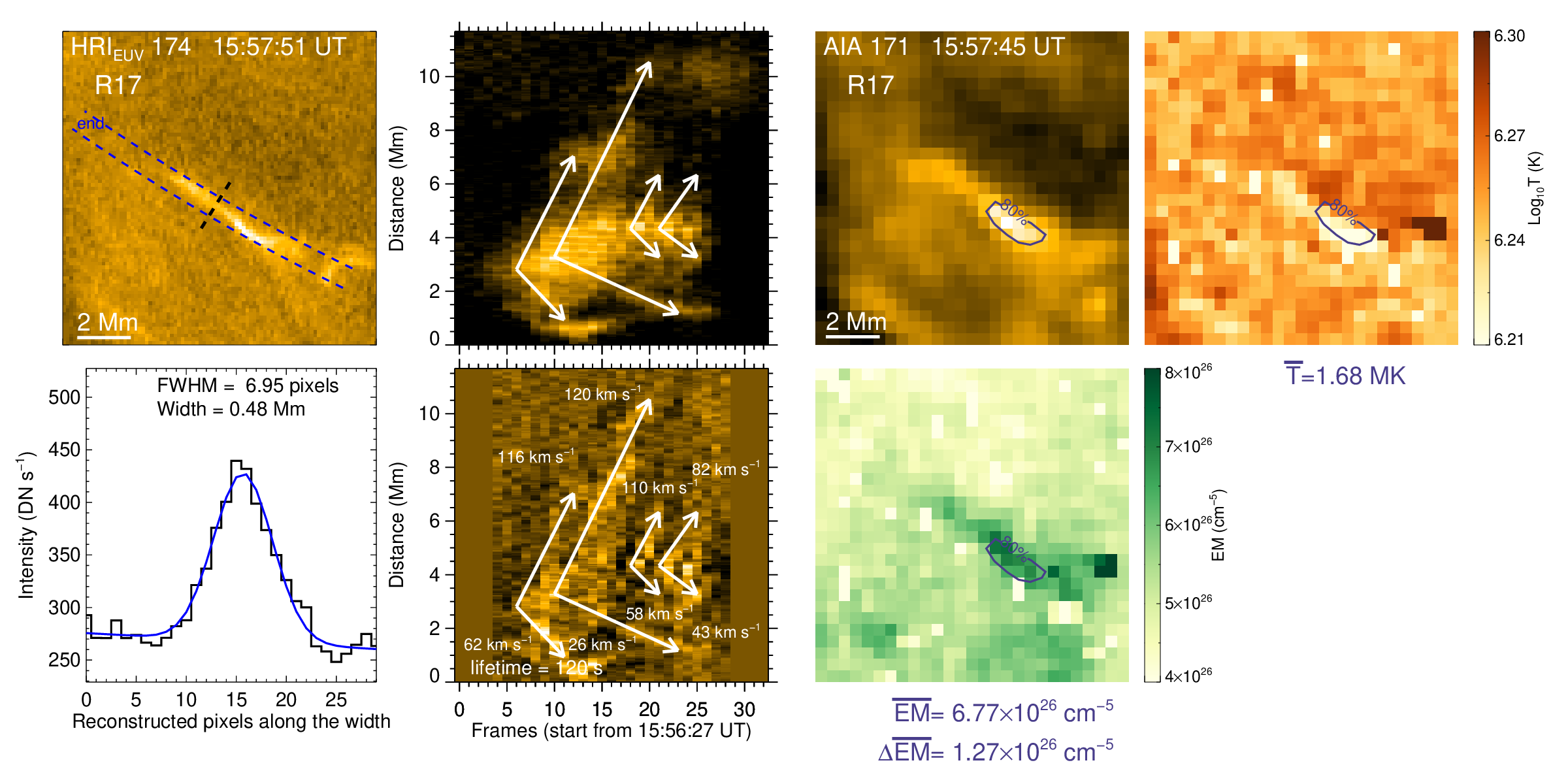}
	}
	\caption{Similar to Figure~\ref{figure3}, the results for R14, R16, and R17.}\label{figure_appenix5}
\end{figure*}

\begin{figure*}[!b]
	\centering
	\subfigure{
		\includegraphics[width=15 cm]{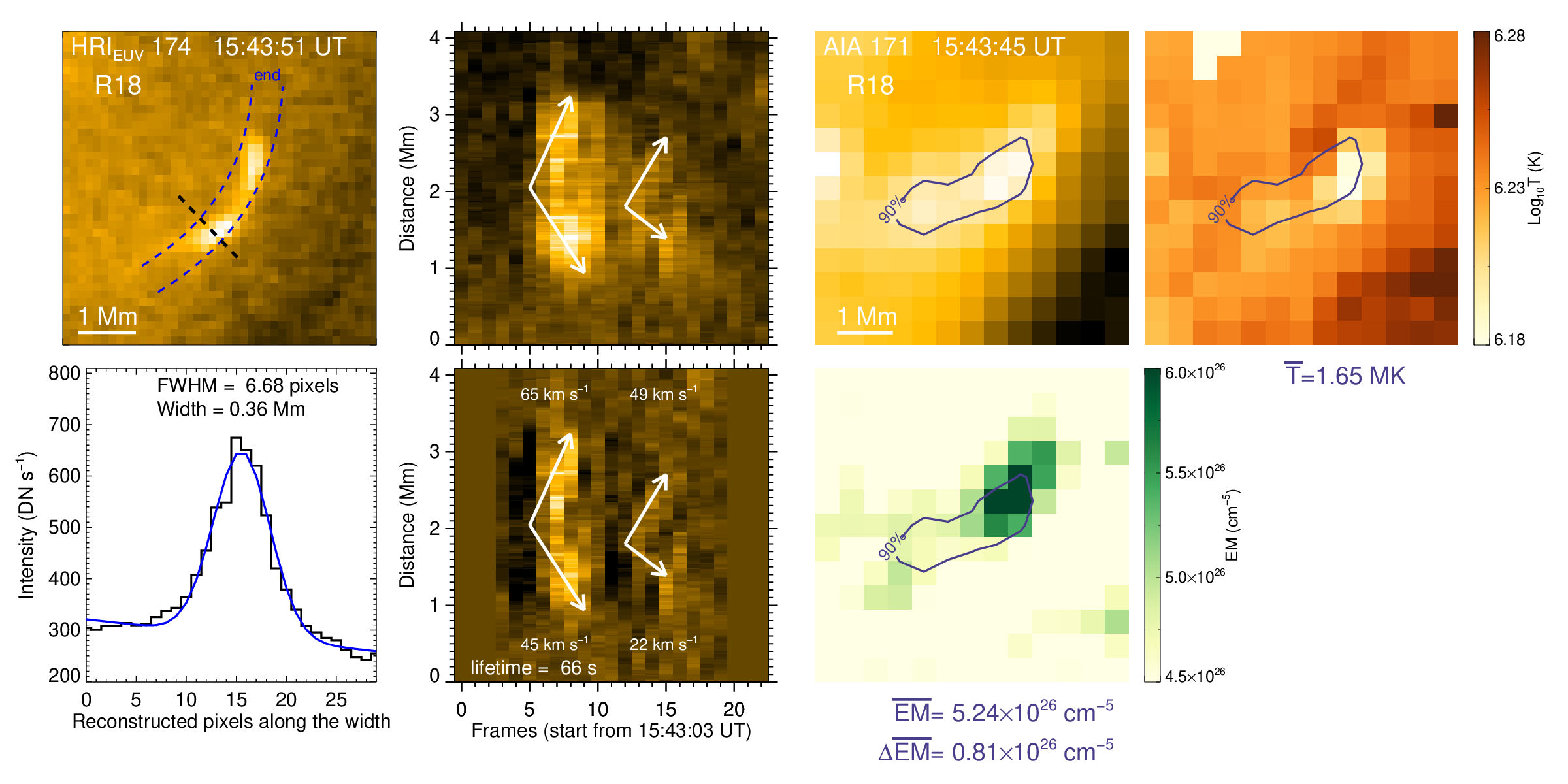}
	}
	\subfigure{
		\includegraphics[width=15 cm]{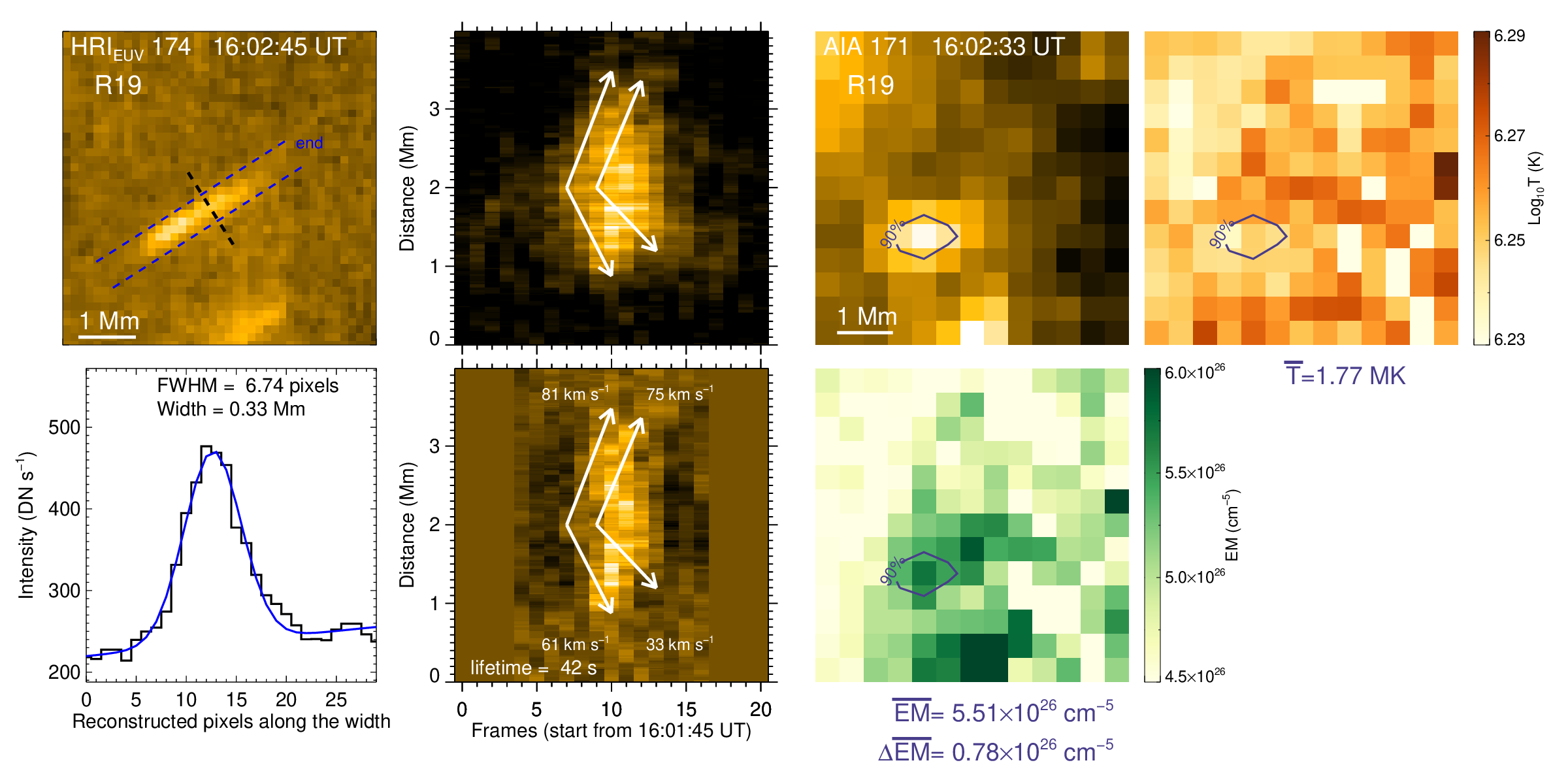}
	}
	\subfigure{
		\includegraphics[width=15 cm]{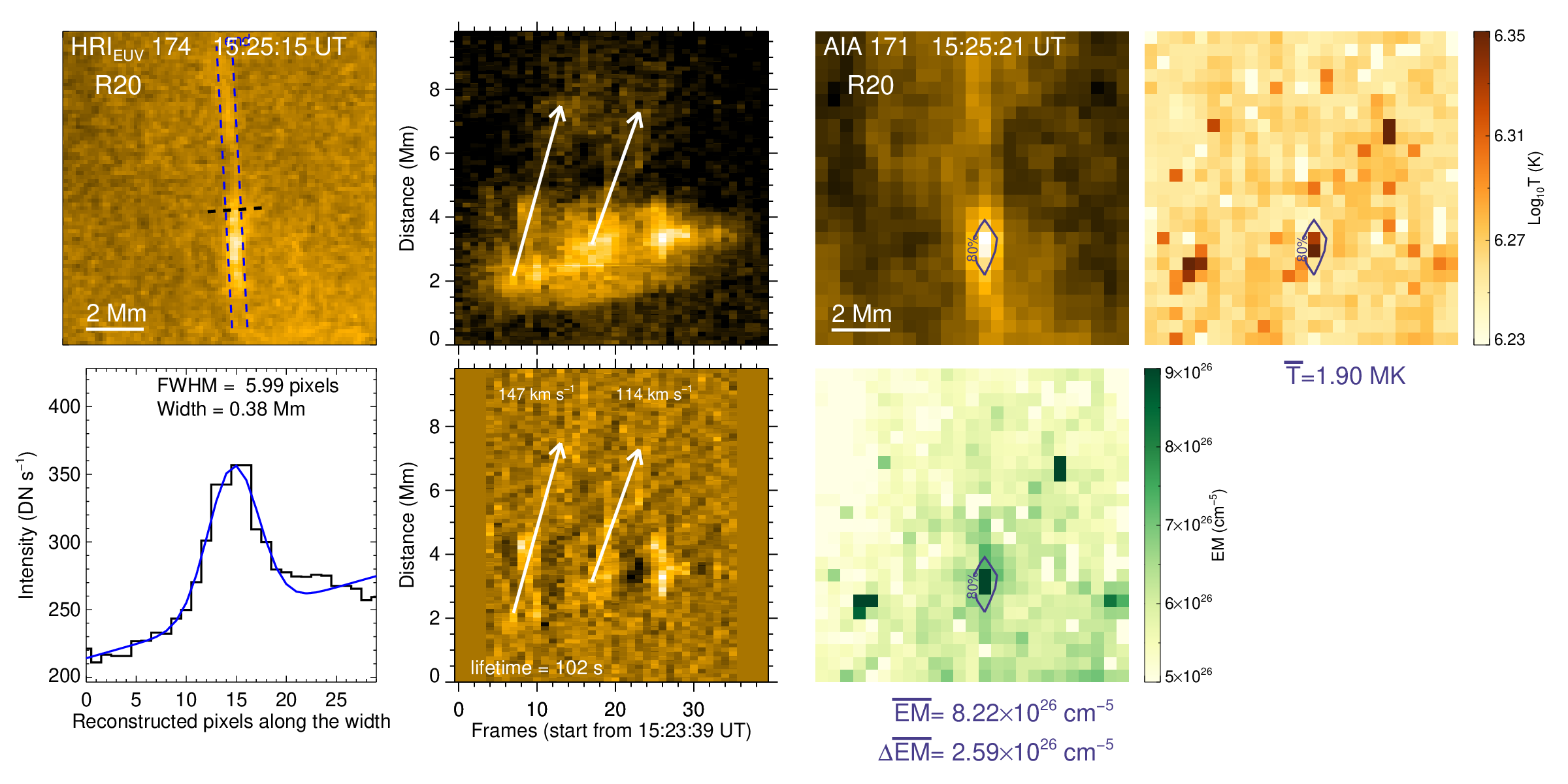}
	}
	\caption{Similar to Figure~\ref{figure3}, the results for R18, R19, and R20.}\label{figure_appenix6}
\end{figure*}

\begin{figure*}[!b]
	\centering
	\subfigure{
		\includegraphics[width=15 cm]{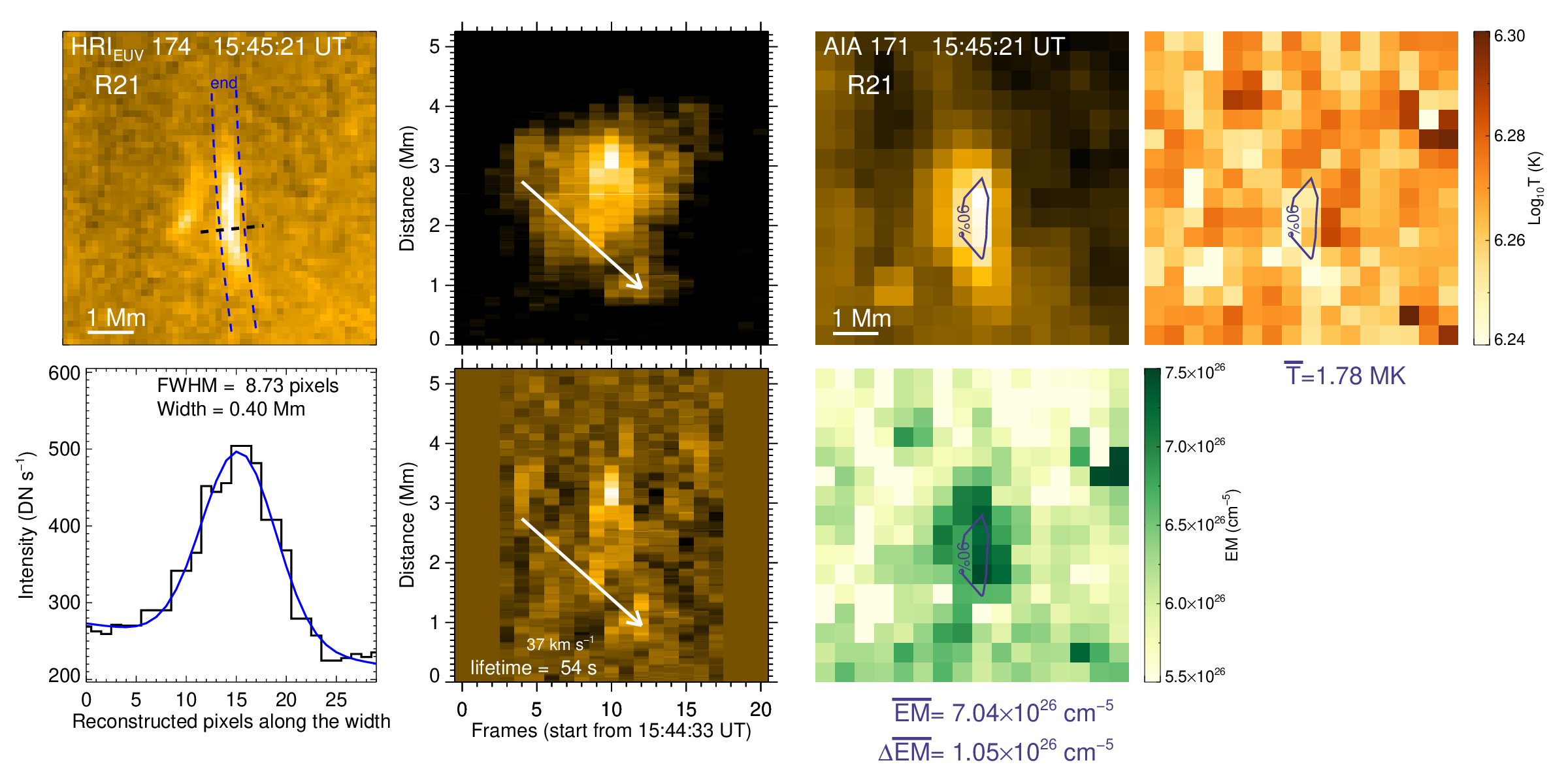}
	}
	\subfigure{
		\includegraphics[width=15 cm]{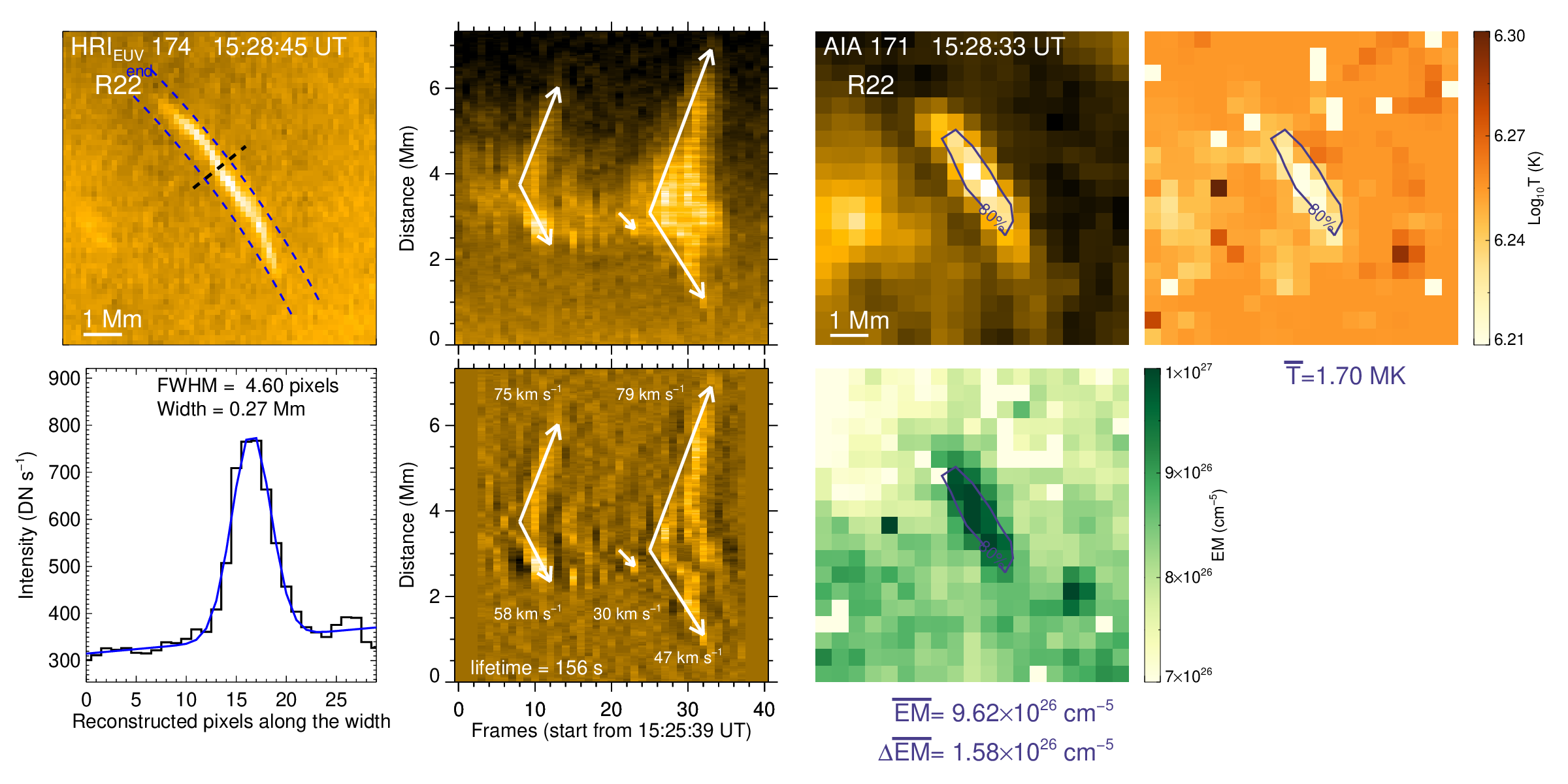}
	}
	\caption{Similar to Figure~\ref{figure3}, the results for R21 and R22.}\label{figure_appenix7}
\end{figure*}

\end{appendix}


\end{document}